\documentclass[
aps,                 
prd,                 
reprint,             
superscriptaddress,  
nofootinbib,         
eprint,              
]{revtex4-2}

\usepackage{graphicx}  
\usepackage{dcolumn}   
\usepackage{bm}        
\usepackage{booktabs}  

\usepackage{amsmath}
\usepackage{amssymb} 
\usepackage{amsthm}

\usepackage{hyperref}
\hypersetup{
    colorlinks=true,    
    linkcolor=blue,     
    citecolor=green,    
    filecolor=magenta,  
    urlcolor=cyan,      
}

\usepackage{cleveref}

\usepackage{glossaries-extra}
\glsdisablehyper  
\newglossaryentry{LCDM}{
    name=$\Lambda$CDM,
    description=$\Lambda$ Cold Dark Matter,
}
\newglossaryentry{SS}{
    name=SS,
    description=standard siren,
    first=standard siren (SS),
    plural=SSs,
    descriptionplural=standard sirens,
    firstplural=Standard Sirens (SSs)
}
\newglossaryentry{SnIa}{
    name=SnIa,
    description=type Ia supernova,
    first=type Ia supernova (SnIa),
    plural=SnIas,
    descriptionplural=type Ia supernovas,
    firstplural=type Ia supernovas (SnIas)
}
\newglossaryentry{CMB}{
    name=CMB,
    description=cosmic microwave background,
    first=cosmic microwave background (CMB),
}
\newglossaryentry{FLRW}{
    name=FLRW,
    description=Friedmann Lemaître Robertson Walker,
    first=FLRW,
}
\newglossaryentry{GW}{
    name=GW,
    description=gravitational wave,
    first=gravitational wave (GW),
    plural=GWs,
    descriptionplural=gravitational waves,
    firstplural=gravitational waves (GWs)
}
\newglossaryentry{GR}{
    name=GR,
    description=General Relativity,
    first=General Relativity (GR)
}
\newglossaryentry{CDM}{
    name=CDM,
    description=cold dark matter,
    first=cold dark matter (CDM)
}
\newglossaryentry{PSIS-LOO-CV}{
    name=PSIS-LOO-CV,
    description=Pareto smoothed importance sampling leave-one-out cross-validation,
    first=Pareto smoothed importance sampling leave-one-out cross-validation (PSIS-LOO-CV)
}
\newglossaryentry{MCMC}{
    name=MCMC,
    description=Markov chain Monte Carlo,
    first=Markov chain Monte Carlo (MCMC),
}
\newglossaryentry{EM}{
    name=EM,
    description=electromagnetic,
    first=electromagnetic (EM)
}
\newglossaryentry{LISA}{
    name=LISA,
    description=Laser Interferometer Space Antenna,
    first=Laser Interferometer Space Antenna (LISA)
}
\newglossaryentry{ET}{
    name=ET,
    description=Einstein Telescope,
    first=Einstein Telescope (ET)
}
\newglossaryentry{LIGO}{
    name=LIGO,
    description=Laser Interferometer Gravitational-Wave Observatory,
    first=Laser Interferometer Gravitational-Wave Observatory (LIGO)
}

\makeatletter
\def\p@subsection{}
\makeatother

\begin{document}

\title{Testing $\Lambda$-Free $f(Q)$ Cosmology}


\author{Jos\'e Ferreira}
\email{jpmferreira@ciencias.ulisboa.pt}
\affiliation{Instituto de Astrof\'isica e Ci\^encias do Espa\c{c}o, Faculdade de Ci\^encias da Universidade de Lisboa, Campo Grande, Edificio C8, P-1749-016, Lisboa, Portugal}
\affiliation{Departamento de F\'isica, Faculdade de Ci\^encias da Universidade de Lisboa, Campo Grande, Edificio C8, P-1749-016, Lisboa, Portugal}

\author{Tiago Barreiro}
\email{tmbarreiro@ulusofona.pt}
\affiliation{Instituto de Astrof\'isica e Ci\^encias do Espa\c{c}o, Faculdade de Ci\^encias da Universidade de Lisboa, Campo Grande, Edificio C8, P-1749-016, Lisboa, Portugal}
\affiliation{ECEO, Universidade Lus\'ofona de Humanidades e Tecnologias, Campo Grande, 376, 1749-024 Lisboa, Portugal}

\author{Jos\'e P. Mimoso}
\email{jpmimoso@ciencias.ulisboa.pt}
\affiliation{Instituto de Astrof\'isica e Ci\^encias do Espa\c{c}o, Faculdade de Ci\^encias da Universidade de Lisboa, Campo Grande, Edificio C8, P-1749-016, Lisboa, Portugal}
\affiliation{Departamento de F\'isica, Faculdade de Ci\^encias da Universidade de Lisboa, Campo Grande, Edificio C8, P-1749-016, Lisboa, Portugal}

\author{Nelson J. Nunes}
\email{njnunes@ciencias.ulisboa.pt}
\affiliation{Instituto de Astrof\'isica e Ci\^encias do Espa\c{c}o, Faculdade de Ci\^encias da Universidade de Lisboa, Campo Grande, Edificio C8, P-1749-016, Lisboa, Portugal}
\affiliation{Departamento de F\'isica, Faculdade de Ci\^encias da Universidade de Lisboa, Campo Grande, Edificio C8, P-1749-016, Lisboa, Portugal}

\date{\today}

\glsunset{LIGO}
\glsunset{LISA}
\glsunset{LCDM}

\begin{abstract}
We study a model of Symmetric Teleparallel gravity that is able to account for the current accelerated expansion of the universe without the need for dark energy component. We investigate this model by making use of dynamical system analysis techniques to identify the regions of the parameter space with viable cosmologies and constrain it using \gls{SnIa}, \gls{CMB} data and make forecasts using \gls{SS} events. We conclude that this model is disfavored with respect to $\Lambda$CDM and forthcoming standard siren events can be decisive in testing the viability of the model.
\end{abstract}

\maketitle


\glsresetall  

\section{Introduction}
\label{sec:introduction}

The standard model of cosmology, referred to as \gls{LCDM}, provides us with the most accurate description of the Universe at large scales. However, despite its successes, it is not devoid of difficulties \cite{LCDM-R2021}. One of the most debated of its shortcomings is the origin of the cosmological constant, introduced to explain the current accelerated expansion of the Universe. Also, tensions in the value of the expansion rate obtained from low redshift against high redshift data, motivate the search for alternative descriptions.

Recently, it has been shown that two gravitational theories equivalent to \gls{GR} exist, one based solely on non-metricity and the other based solely on torsion. This equivalence is referred as the geometrical trinity of gravity \cite{Jimenez2019a}. Generalization of these formulations have been considered as $f(T)$, the teleparallel gravity, or $f(Q)$, the symmetric teleparallel gravity \cite{Jimenez2017,Cai2016,MGCosmology2021}. These theories are particularly interesting as even if they only produce slight deviations from \gls{LCDM} at the background level \cite{Ayuso2021}, they may however, give rise to signatures in the evolution of the density perturbations \cite{Barros2020} and in the propagation of gravitational waves \cite{Belgacem2017a}, potentially being able to solve the challenges that \gls{LCDM} faces \cite{noemi1,noemi2,noemi3,noemi4,defalco,colgain,andronikos,Ayuso2020}.

With the development of new techniques in \gls{GW} astronomy, cosmologists will soon have access to a new and exciting source of data: \gls{SS} events, characterized by the direct measurement of both the luminosity distance and the redshift of merging compact objects. Unfortunately, up to the current date, only GW170817 \cite{GW170817} has been confirmed as a \gls{SS} event and a suggested \gls{EM} counterpart to GW190521 \cite{GW190521} was proposed in \cite{GW190521-EM}. 

In this work, we focus on a model of $f(Q)$ gravity able to account for the late time accelerated expansion of the Universe without adding a cosmological constant. We develop a dynamical system analysis in order to look for regions in parameter space that yield viable cosmologies, and constrain this model by making use of both high and low redshift observables, namely, \gls{SnIa}, the \gls{CMB} and in addition, make forecasts using mock catalogs of \gls{SS} events made for \gls{LISA} \cite{LISA-proposal,Auclair2022} and the \gls{ET}. An online repository complementary to this analysis is publicly available at \cite{repo}.

This work is organized as follows: In \cref{sec:model} we provide a brief introduction to the formalism behind $f(Q)$ cosmology and introduce the specific model we will be working with. In \cref{sec:dynamical-system}, we perform a dynamical system analysis applied to the model to identify the regions of the phase space leading to late time acceleration. In \cref{sec:datasets} we introduce the datasets we use.
In \cref{sec:constraints} we show the constraints given by the datasets introduced before. Finally, in \cref{sec:conclusions}, we make an overview of this work.

\section{Cosmology in $f(Q)$}
\label{sec:model}

In this work, we consider a theory of modified gravity based on an arbitrary function of the non-metricity scalar. The action for such a theory of gravity reads
\begin{equation}
    \label{eq:action}
    S = \int \sqrt{-g} \left[ -\frac{c^4}{16 \pi G} f(Q) + \mathcal{L}_m \right] d^4x \,,
\end{equation}
where $Q$ is the non-metricity scalar that is defined as \cite{Jaerv2018}
\begin{equation}
    \label{eq:Q}
    Q = -\frac{1}{4} Q_{\alpha \beta \gamma} Q^{\alpha \beta \gamma} +
    \frac{1}{2} Q_{\alpha \beta \gamma} Q^{ \gamma \beta \alpha} + 
    \frac{1}{4} Q_{\alpha}Q^{\alpha} 
    - \frac{1}{2} Q_{\alpha} \tilde{Q}^{\alpha} \,,
\end{equation}
with $Q_{\alpha\mu\nu} = \nabla_\alpha \, g_{\mu\nu}$ and the two independent contractions of the non-metricity tensor are
\begin{equation}
    Q_\mu = Q_\mu{}^\alpha{}_\alpha \,, \hspace{1cm}
    \tilde{Q}^\mu = Q_\alpha{}^{\alpha \mu} \,.
\end{equation}
When varying the action in \cref{eq:action} with respect to the metric, one obtains the field equations \cite{Jimenez2019}
\begin{eqnarray}
    \label{eq:field-equations}
    \frac{2}{\sqrt{-g}} \nabla_\alpha \left( \sqrt{-g} f_{Q} P^{\alpha\mu}{}_\nu \right) 
    +  \frac{1}{2} \delta^\mu_\nu f  \nonumber \\
    +  f_{Q} P^{\mu\alpha\beta} Q_{\nu\alpha\beta}   
    =  \frac{8 \pi G}{c^4} T^\mu{}_\nu \,,
\end{eqnarray}
where $f_Q$ represents the partial derivative of $f(Q)$ with respect to $Q$. 
$P^{\alpha \mu \nu}$, referred to as the non-metricity conjugate, is defined as
\begin{equation}
    P^\alpha{}_{\mu \nu} = - \frac{1}{2} L^\alpha{}_{\mu \nu} + \frac{1}{4} (Q^\alpha - \tilde{Q}^\alpha) - \frac{1}{4} \delta^\alpha_{(\mu} Q_{\nu)} \,.
\end{equation}
where $L^\alpha{}_{\mu \nu}$ is the disformation tensor
\begin{equation}
    L^{\lambda}{}_{\mu\nu} = \frac{1}{2} g^{\lambda \beta} 
    \left( -Q_{\mu \beta\nu}-Q_{\nu \beta\mu}+Q_{\beta \mu \nu} \right) \,.
\end{equation}
We use the coincident gauge to specify the affine connection \cite{Jimenez2017,Jimenez2018,Jaerv2018} 
and restrict our analysis to a flat, homogeneous and isotropic universe, described by the \gls{FLRW} metric
\begin{equation}
    ds^2 = -c^2dt^2 + a^2(t)(dx^2 + dy^2 + dz^2) \,,
\end{equation}
where $a(t)$ is the scale factor of the universe, and $t$ is cosmic time $t$. Under these assumptions, the non-metricity scalar defined in \cref{eq:Q} simplifies to \cite{Jimenez2019}
\begin{equation}
    \label{eq:Q=6H2}
    Q = 6H^2 \,,
\end{equation}
where $H = \dot{a}/a$ is the Hubble function.
Considering a universe permeated by a perfect fluid composed of both matter and radiation, the modified first Friedmann equation reads \cite{Jimenez2019}
\begin{equation}
    \label{eq:friedmann}
    6 f_Q H^2 - \frac{1}{2}f = 8 \pi G (\rho_m + \rho_r) \,.
\end{equation}
As stated in \cite{Koivisto2005}, the covariant derivative of the stress-energy momentum tensor is still zero, meaning that each component of the fluid obeys the standard continuity equation
\begin{equation}
    \label{eq:continuity}
    \dot{\rho} + 3 \frac{\dot{a}}{a} \left(\rho + \frac{P}{c^2} \right) = 0 \,.
\end{equation}
where $P$ stands for pressure. This means that the scaling for each component is the same as in $\Lambda$CDM and we can rewrite \cref{eq:friedmann} as
\begin{equation}
    \label{eq:friedmann-w-omegas}
    \frac{2Qf_Q - f}{Q_0} = \frac{\Omega_m}{a^3} + \frac{\Omega_r}{a^4} \,,
\end{equation}
where $\Omega_i = 8 \pi G\rho_{i,0}/3H_0^2$ and $E \equiv H/H_0$, with the index $0$ denoting the value of that quantity today.

This modification of gravity also changes the equation of motion for the propagation of gravitational waves \cite{Jimenez2019}
\begin{equation}
    \Bar{h}_A'' + 2 \mathcal{H} (1 + \delta(z)) \Bar{h}_A' + k^2 \Bar{h}_A = 0 \,,
\end{equation}
where $A = \times, +$ represent the polarizations of the \gls{GW}.
The previous equation differs from the case of \gls{LCDM} by an additional friction factor, $\delta(z)$, that for an $f(Q)$ model takes the form
\begin{equation}
    \delta(z) = \frac{d \ln{f_Q}}{2 \mathcal{H} d\eta} \,.
\end{equation}
Following \cite{Belgacem2017a}, this leads to a modification to the luminosity distance for a \gls{GW} event, such that
\begin{equation}
    \label{eq:dlGW}
    d_\text{GW} (z) = \sqrt{\frac{f_Q^{(0)}}{f_Q}} \,\, d_L(z) \,,
\end{equation}
where $f_Q^{(0)}$ is the function $f_Q$ computed at the present day and $d_L(z)$ is the standard luminosity distance function.

\subsection{$f(Q)$ Cosmological Model}

In this work we consider a model that differs from the standard interpretation of gravity, i.e., \gls{GR}, for which $f(Q) = Q$, by taking a multiplicative exponential term of the form \cite{Anagnostopoulos2021},
\begin{equation}
    \label{eq:model}
    f(Q) = Qe^{\lambda Q_0 /Q} \,,
\end{equation}
where $\lambda$ is a constant, however not a free parameter. As we will see in \cref{sec:dynamical-system}, $\lambda$ is related to the abundance of matter today and has to be positive for a viable universe.

By making use of \cref{eq:Q=6H2}, we can see that the extra multiplicative term in this model is an exponential that is inversely proportional to the Hubble function. Given that the Hubble function increase with redshift, this model exponentially approaches the case where $f(Q) = Q$, meaning that it has \gls{GR} as a limit case at early times.

Inserting the specific form of $f(Q)$ into \cref{eq:friedmann-w-omegas} we obtain the modified first Friedmann equation for this model, which reads
\begin{equation}
    \label{eq:E-model}
    (E^2 - 2\lambda)e^{\lambda/E^2} = \Omega_m (1+z)^3 + \Omega_r (1+z)^4 \,,
\end{equation}
where $E(z) \equiv H(z)/H_0$.

As hinted before, this model falls back to \gls{GR} in the limit of high redshifts. For $z \gg 1$, \cref{eq:E-model} is approximately equal to
\begin{equation}
    \label{eq:E-model-highredshift}
    E^2 \approx \Omega_m (1+z)^3 + \Omega_r (1+z)^4 + \lambda \,,
\end{equation}
which corresponds to a \gls{LCDM} universe with $\Omega_\Lambda = \lambda$, where its effect is negligible at these early times. When evaluated today $(E^2 \approx 1)$, \cref{eq:E-model} reduces to
\begin{equation}
    \label{eq:E-model-lowredshift}
    E^2 \approx e^{-\lambda} \Omega_m (1+z)^3 + e^{-\lambda} \Omega_r (1+z)^4 + 2\lambda \,,
\end{equation}
which corresponds once again to a \gls{LCDM} universe, but now with $\Omega_\Lambda = 2 \lambda$ and a correction in the relative abundance of matter and radiation. We note that, in the absence of matter fields, $\rho_m = \rho_r = 0$, we obtain a de Sitter universe with $\lambda = 1/2$. 

If we insert the form of the function $f(Q)$ in \cref{eq:dlGW}, then we obtain the luminosity distance function for this specific model which now reads
\begin{equation}
    \label{eq:dlGW-model}
    d_\text{GW}(z) = \sqrt{ \frac{1 - \lambda}{1 - \lambda/E^2}} e^{\frac{\lambda}{2}(1 - 1/E^2)} d_L(z) \,,
\end{equation}

This model was originally proposed in \cite{Anagnostopoulos2021}, where it was shown to be statistically equivalent to \gls{LCDM} for tests using low redshift measurements. Additionally, it has been shown to pass Big Bang nucleosynthesis constraints \cite{Anagnostopoulos2022}.

\section{Dynamical System Analysis}
\label{sec:dynamical-system}

In order to perform our dynamical systems analysis, we can start by re-writing \cref{eq:E-model} as
\begin{equation}
    1 = \frac{e^{- \lambda / E^2}}{E^2} \left( \frac{\Omega_m}{a^3} 
    + \frac{\Omega_r}{a^4} \right) + \frac{2\lambda}{E^2} \,.
\end{equation}
Defining the quantities
\begin{gather}
    x_1 \equiv \frac{\Omega_m}{E^2 e^{\lambda / E^2} a^3} \,, \hspace{0.25cm}
    x_2 \equiv \frac{\Omega_r}{E^2 e^{\lambda / E^2} a^4} \,, \hspace{0.25cm}
    x_3 \equiv \frac{2 \lambda}{E^2} \,,
\end{gather}
where $x_1$ is related to the evolution of the matter density in the universe, $x_2$ with the radiation density and $x_3$ with the parameter $\lambda$, they satisfy
\begin{equation}
    \label{eq:dynamical-system-conservation}
    x_1 + x_2 + x_3 = 1 \,.
\end{equation}
This equation shows that one of our coordinates in phase space is fully determined by the value of the other two. As such, without loss of generality, we choose to work with $x_1$ and $x_2$ alone.

Differentiating both $x_1$ and $x_2$ with respect to the number of $e$-folds, $N = \ln{a}$, we obtain
\begin{align}    
    \label{eq:x1prime}
    x_1' = - x_1 \left((1+x_1+x_2)\frac{E'}{E} + 3\right) \,, \\
    \label{eq:x2prime}
    x_2' = - x_2 \left((1+x_1+x_2)\frac{E'}{E} + 4\right) \,,
\end{align}
where the prime denotes the derivative with respect to $N$ and the ratio $E'/E$ can be expressed as a function of both $x_1$ and $x_2$ by
\begin{equation}
    \frac{E'}{E} = \frac{ -(3 x_1 + 4 x_2)}{2 - (x_1 + x_2) + (x_1 + x_2)^2} \,.
\end{equation}

\subsection{Fixed Points}

For this dynamical system, we find three distinct fixed points. Their locations and corresponding stability is presented in \cref{tab:fixed-points}.

\begin{table}[h!]
    \centering
    \begin{tabular}{|c|c|c|c|c|c|c|}
        \hline
        Point & Type                 & $x_1$ & $x_2$ & Stability & Eigenvalues \\
        \hline
        \textit{I}  & $\lambda$ dominated  & 0     & 0     & Stable    & $(-4, -3)$    \\
        \hline
        \textit{II} & Matter dominated     & 1     & 0     & Saddle    & $(3, -1)$     \\
        \hline
        \textit{III} & Radiation dominated & 0     & 1     & Unstable  & $(4, 1)$      \\
        \hline
    \end{tabular}
    \caption{Fixed points and corresponding stability for the dynamical system.}
    \label{tab:fixed-points}
\end{table}

Fixed point \textit{I} corresponds to a $\lambda$ dominated regime and is the only stable fixed point in our system. Located at $x_1,x_2 = 0$, corresponding to
\begin{equation}
    E^2 = 2 \lambda \,,
\end{equation}
and therefore it only exists if $\lambda$ is positive.

By integrating the previous equation we obtain the evolution of the scale factor as a function of time
\begin{equation}
    a = e^{\sqrt{2 \lambda} H_0 t} \,,
\end{equation}
meaning that we have an exponential expansion, corresponding to a point of eternal inflation.

The fixed point \textit{II} corresponds to a matter dominated regime located at $x_1 = 1$ and $x_2 = 0$, implying that $\lambda = 0$. It is a saddle node with attracting trajectories along the direction of fixed point \textit{III}. Setting $\lambda = 0$ in \cref{eq:model}, we have $f(Q) = Q$, corresponding to the \gls{GR} limit, more specifically a \gls{CDM} universe, since this model does not include a cosmological constant.

As for fixed point \textit{III}, it corresponds to a radiation dominated regime and is located at $x_1 = 0$ and $x_2 = 1$. It is an unstable node and like fixed point \textit{II} it is located in a region where $\lambda = 0$, meaning that it is in the \gls{GR} limit with $f(Q) = Q$.

\subsection{Trajectories}
\begin{figure}[h!]
    \centering
    \includegraphics[width=\columnwidth]{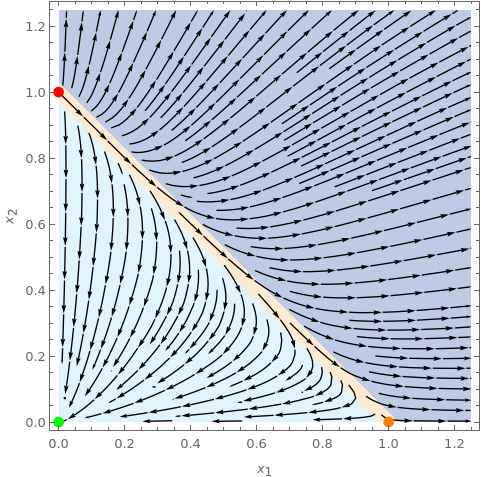}
    \caption{Stream plot of the phase space for the dynamical system where fixed point \textit{I} (stable) is represented in green, fixed point \textit{II} (saddle) in orange and fixed point \textit{III} (unstable) in red. The light blue marks the region where $\lambda > 0$, the light brown region corresponds to $\lambda = 0$ and the dark blue region to $\lambda < 0$.}
    \label{fig:dynamical-system}
\end{figure}

Solving the equations of motion \cref{eq:x1prime,eq:x2prime} numerically, we obtain the trajectories in phase space presented in \cref{fig:dynamical-system}. From this figure, we can see the existence of three disjoint regions related with the sign of $\lambda$:
\begin{itemize}
    \item $\lambda > 0$: region colored in light blue, corresponding to the triangle $x_1 + x_2 < 1$. This is where the stable fixed point \textit{I} is located, therefore leading all trajectories inside this region towards a universe in eternal inflation.
    
    \item $\lambda = 0$: region in light brown, corresponding to the line $x_1 + x_2 = 1$ and a \gls{CDM} universe. In this region the trajectories flow from the unstable fixed point \textit{III}, corresponding to a radiation dominated epoch, towards the saddle fixed point \textit{II}, corresponding to a matter dominated epoch.
    
    \item $\lambda < 0$: region in dark blue, corresponding to the plane $x_1 + x_2 > 1$. It has no fixed points and all trajectories diverge towards infinity, with $E$ asymptotically approaching zero.
\end{itemize}

In order to agree with the observations of an expanding universe the value of $\lambda$ must be positive. This is what was obtained in \cite{Anagnostopoulos2021} where this model was constrained using \gls{SnIa}, baryonic acoustic oscillations, cosmic chronometers, and redshift space distortions.

Qualitatively, the previous dynamical system is similar to a \gls{LCDM} universe where the value of the cosmological constant can be positive, negative or zero \cite{Bahamonde2018}. However, in \gls{LCDM} with a negative cosmological constant, the universe eventually collapses into a Big Crunch.

\subsection{Computing the value of $\lambda$}
In order to compute the value of $\lambda$, we evaluate \cref{eq:E-model} at the present day, obtaining
\begin{equation}
    \label{eq:friedmann-present}
    \left( \lambda - \frac{1}{2} \right) e^{\lambda - 1/2} = - \frac{\Omega_m + \Omega_r}{2\sqrt{e}} \,.
\end{equation}
This equation only possesses a solution provided that $\Omega_m + \Omega_r \leq 2 e^{-1/2}$. This is not immediately ensured because the sum of the densities is not necessarily equal to 1. The two  possible solutions for $\lambda$ are
\begin{eqnarray}
    \lambda_0 &=& \frac{1}{2} + W_0\left( -\frac{\Omega_m + \Omega_r}{2e^{1/2}} \right) \,, \\
    \lambda_{-1} &=& \frac{1}{2} + W_{-1}\left( -\frac{\Omega_m + \Omega_r}{2e^{1/2}} \right) \,,
\end{eqnarray}
where $W_0$ and $W_{-1}$ are the main and the $-1$ branch of the Lambert function, respectively.

We discard the solution $\lambda_{-1}$, since it only provides negative values for $\lambda$. Additionally, we see that $\lambda_0$ takes negative values when $\Omega_m + \Omega_r > 1$, hence, for $\lambda$ to be positive, the sum of these densities must be smaller or equal to 1 as in \gls{LCDM}.

\section{Datasets}
\label{sec:datasets}

Throughout this work we make use of two different sources of observational data, \gls{SnIa} and \gls{CMB}, and of mock catalogs for \gls{SS} events generated for the \gls{LISA} space mission and \gls{ET}.

\subsection{Type Ia Supernovae}

We use \gls{SnIa} to provide low redshift constraints for this model. The dataset of \gls{SnIa} used is the Pantheon sample \cite{pantheon}, with the corresponding data publicly available in \cite{pantheon-repo}. For performance reasons, the binned sample of this dataset was used.

The relationship between the apparent magnitude and the luminosity distance is given by
\begin{equation}
    \label{eq:m1}
    m = M + 5 \log{\left( \frac{d_L(z)}{\rm Mpc} \right)} + 25 \,,
\end{equation}
where $M$ is the bolometric magnitude. In order to avoid degeneracies and unwanted parameters, we followed the procedure presented in \cite{Goliath2001}, where a marginalization was performed on both $H_0$ and the bolometric magnitude. The likelihood for the \gls{SnIa} reads
\begin{equation}
    \label{eq:likelihood-snia}
    L = \exp \left[ -\frac{1}{2} \left( A - \frac{B^2}{C} \right)  \right] \,,
\end{equation}
where $A$, $B$ and $C$ are defined as
\begin{gather}
    A = \sum_{i = 1}^n \frac{\Delta^2(z_i)}{\sigma^2(z_i)} \,, 
    B = \sum_{i = 1}^n \frac{\Delta(z_i)}{\sigma^2(z_i)} \,,
    C = \sum_{i = 1}^n \frac{1}{\sigma^2(z_i)} \,,
\end{gather}
with $\sigma(z)$ representing the error for each \gls{SnIa} measurement at redshift $z_i$ and where
\begin{equation}
    \Delta (z_i) = m^{\text{(obs)}}(z_i) - 5\log{\left( \frac{H_0}{c}d_L(z_i) \right)} \,,
\end{equation}
is the difference between the observed magnitude $m^{\text{(obs)}}$ and the $H_0$-independent luminosity distance.

\subsection{Cosmic Microwave Background Radiation}

In order to assess the behavior of this model at high redshifts, we condense the information provided by the \gls{CMB} into what are known as the shift parameters \cite{Zhai2018}
\begin{eqnarray}
    R        &=& \sqrt{\Omega_b + \Omega_c}\, \frac{H_0}{c}\, r(z_*) \,, \\
    l_a      &=& \pi \frac{r(z_*)}{r_s(z_*)} \,, \\
    \omega_b &=& \Omega_b h^2 \,,
\end{eqnarray}
where $\Omega_m = \Omega_b + \Omega_c$ is the sum of the baryonic and dark matter densities, $r(z)$ is the comoving distance 
\begin{equation}
    r(z) = \int_0^z \frac{c}{H(z)} dz \,,
\end{equation}
and $r_s(z)$ is the comoving sound horizon
\begin{equation}
    r_s(z) = \int_z^\infty \frac{c_s(z)}{H(z)} dz \,,
\end{equation}
with $c_s(z)$, the sound speed as a function of redshift
\begin{equation}
    c_s(z) = \frac{c}{\sqrt{3(1+\hat{R}_b/(1+z))}} \,,
\end{equation}
where $\hat{R}_b$ is given by
\begin{equation}
    \hat{R}_b = 31500\, \Omega_b h^2 \left( \frac{T_{\rm CMB}}{2.7 K} \right)^{-4} \,,
\end{equation}
with the temperature of the \gls{CMB} being taken as $T_\text{CMB} = 2.7255\,$K.

We can compute the value of the redshift of photon decoupling surface, $z_*$, as \cite{Hu1995}
\begin{equation}
    z_* = 1048 \left[1 + \frac{0.00124}{(\Omega_b h^2)^{0.738}}\right] \left[1 + g_1(\Omega_m h^2)^{g_2}\right] \,,
\end{equation}
where $g_1$ and $g_2$ are
\begin{eqnarray}
    g_1 &=& \frac{0.0783\,(\Omega_b h^2)^{-0.238}}{1 + 39.5(\Omega_b h^2)^{-0.763}} \,, \\
    g_2 &=& 0.560 \left[1 + 21.1\,(\Omega_b h^2)^{1.81}\right]^{-1} \,.
\end{eqnarray}
Besides the shift parameters, we fix the value of $\omega_r = \Omega_r h^2$ to $\omega_r = 4.15 \times 10^{-5}$. 

We consider a Gaussian likelihood of the form
\begin{equation}
    L = \exp{ \left( -\frac{1}{2} \Vec{x}\,^T C^{-1} \Vec{x} \right)} \,,
\end{equation}
where $\Vec{x} = (l_a - l_a^\text{(exp)}, R - R^\text{(exp)}, \omega_b - \omega_b^\text{(exp)})$ is the vector formed by the theoretical prediction and the measurements of the three quantities defined above. $C^{-1}$ is the inverse of the covariance matrix.  Both the observational values and the covariance matrix were taken from the Planck 2018 TT+lowE release \cite{Perez2023,Planck2018}.

\subsection{Standard Sirens}

Even though this model of $f(Q)$ quickly recovers \gls{GR} as the redshift increases, the \gls{GW} luminosity distance, presented previously in \cref{eq:dlGW-model}, is not identical to the standard luminosity distance function. Due to the lack of constraining power coming from current \gls{SS} events, we instead rely on forecast data in order to assess the potential that current and future \gls{GW} observatories will have in constraining this model.

To generate the mock catalogs of \gls{SS} events we took the procedure presented in \cite{Ferreira2022}, where mock catalogs for \gls{LIGO}, \gls{LISA} and the \gls{ET} were developed. In short, the authors took the theoretical expected distribution of observed \gls{SS} events and the expected error for each measurements for a given observatory, and, assuming a \gls{LCDM} cosmological model in agreement with \gls{SnIa}, generated catalogs of \gls{SS} events consisting of a redshift $z$, the value of the luminosity distance $d_\text{GW}(z)$ and its corresponding error $\sigma(z)$.

For this analysis we considered two distinct cases, an \gls{ET} catalog, with 1000 events, and a \gls{LISA} catalog, with 15 events and a constraining power corresponding to the median of the generated catalogs in the previously mentioned article. We considered mock catalogs for \gls{LIGO} but did not use them here as the events provided do not place substantial constraints.

We consider a standard Gaussian likelihood for this dataset of the form
\begin{equation}
    L = \exp \left(- \frac{1}{2} \sum_{i=0}^N \left[ \frac{d_\text{GW}^{\text{(obs)} }(z_i) - d_{\text{GW}}(z_i)}{\sigma(z_i)} \right]^2 \right) \,,
\end{equation}
where $\sigma(z_i)$ represents the error of the measurement of the \gls{SS} events with redshift $z_i$ with the corresponding observed luminosity distance $d_\text{GW}^\text{(obs)}(z_i)$ and $N$ represents the number of \gls{SS} events in each catalog.
\section{Constraints}
\label{sec:constraints}

\subsection{Bayesian Inference Methodology}
We performed a Bayesian analysis using \gls{MCMC} methods. We use PyStan \cite{PyStan}, a Python interface to Stan \cite{Stan}, a statistical programming language that implements the No-U-turn sampler, a variant of the Hamiltonian Monte Carlo. The output was then analyzed using GetDist \cite{GetDist} and ArviZ \cite{arviz} Python packages, that allow for an exploratory analysis of Bayesian models.

For each run we executed at least four independent chains, each of which with at least 2500 samples on the posterior distribution and at least 2500 warm-up steps. To ensure that the chains converged, we used the $\hat{R}$ diagnostic, a convergence test introduced in \cite{Vehtari2021}. This diagnostic indicates how well the different chains are mixed together and whether they have converged. We have only used samples that have $\hat{R} \leq 1.05$, where $\hat{R} = 1$ represents perfect convergence.

The initial values are randomly sampled from a Gaussian distribution with mean around the expected value and a standard deviation of order 10\% of the corresponding mean for each parameter.

We considered weakly informative priors on each parameter, specifically a Gaussian distribution centered on the expected value with a standard deviation two orders of magnitude larger. This ensures a quasi flat prior around the region of interest.

\subsection{Results and Discussion}

Using the methodology described in the previous subsection, we used the datasets presented in \cref{sec:datasets} to constrain both the $f(Q)$ model under study and \gls{LCDM}.

\begin{figure}[t]
    \centering
    \includegraphics[width=0.9\columnwidth]{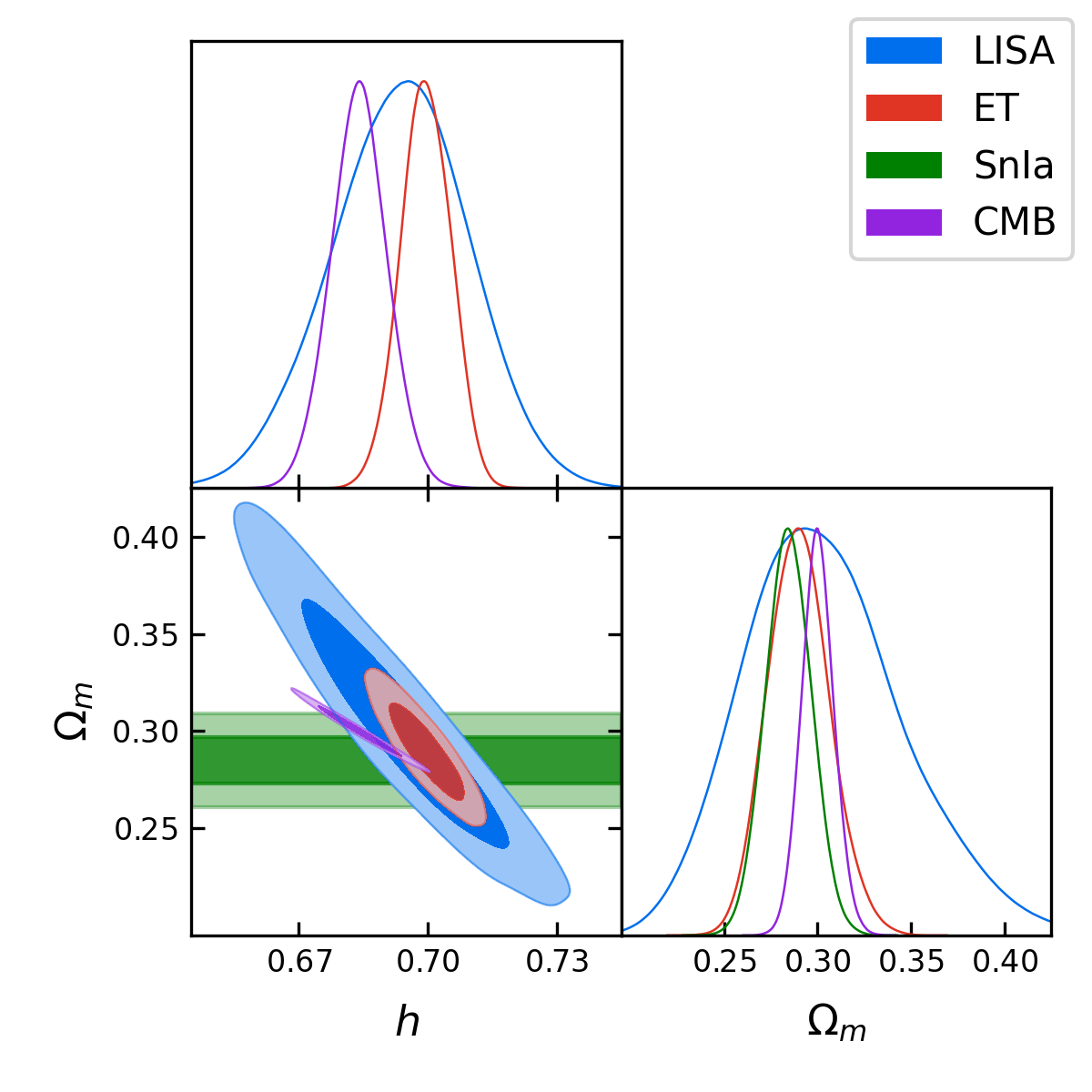}
    \caption{
    Constraints set by \gls{SnIa}, \gls{CMB} shift parameters and forecast \gls{SS} events for the \gls{ET} and a median \gls{LISA} catalog for \gls{LCDM}.
    \vspace{5pt}  
    }
    \label{fig:LCDM-ET-LISA-CMB-SnIa}
\end{figure}

Starting with the constraints on \gls{LCDM}, the corner plot is presented in \cref{fig:LCDM-ET-LISA-CMB-SnIa}, and the best fit values with the corresponding 1$\sigma$ region are presented in \cref{tab:LCDM-ET-LISA-CMB-SnIa}.

\begin{table}[t]
    \begin{tabular}{ccccc}
        \hline                                                                                             \\[-5pt]
                   & SnIa              & CMB                       & LISA              & ET                \\[4pt]
        \hline\hline                                                                                       \\[-5pt]
        $\Omega_m$ & $0.285 \pm 0.012$ & $0.300 \pm 0.009$ & $0.304^{+0.035}_{-0.048}$ & $0.290 \pm 0.017$ \\[2.5pt]
        $h$        & \rule{1cm}{0.1mm} & $0.684 \pm 0.007$ & $0.694 \pm 0.016$         & $0.699 \pm 0.006$ \\[4pt]
        \hline
    \end{tabular}
    \caption{Best fit and 1$\sigma$ error set by \gls{SnIa}, \gls{CMB} and forecast \gls{SS} events for \gls{LISA} and the \gls{ET}, for \gls{LCDM}.
    \vspace{12.5pt}  
    }
    \label{tab:LCDM-ET-LISA-CMB-SnIa}
\end{table}

From \cref{fig:LCDM-ET-LISA-CMB-SnIa}, we can see that in \gls{LCDM} there is an agreement in the value of $\Omega_m$ for all datasets, with \gls{LISA} showing higher error bars when compared to the other datasets. As for the value of $h$ there is a slight disagreement between the \gls{CMB} and the \gls{SS} mock catalogs, given that the mock catalogs used, constructed in \cite{Ferreira2022}, had $h = 0.7$ as fiducial value.

\begin{figure}[t]
    \centering
    \includegraphics[width=0.9\columnwidth]{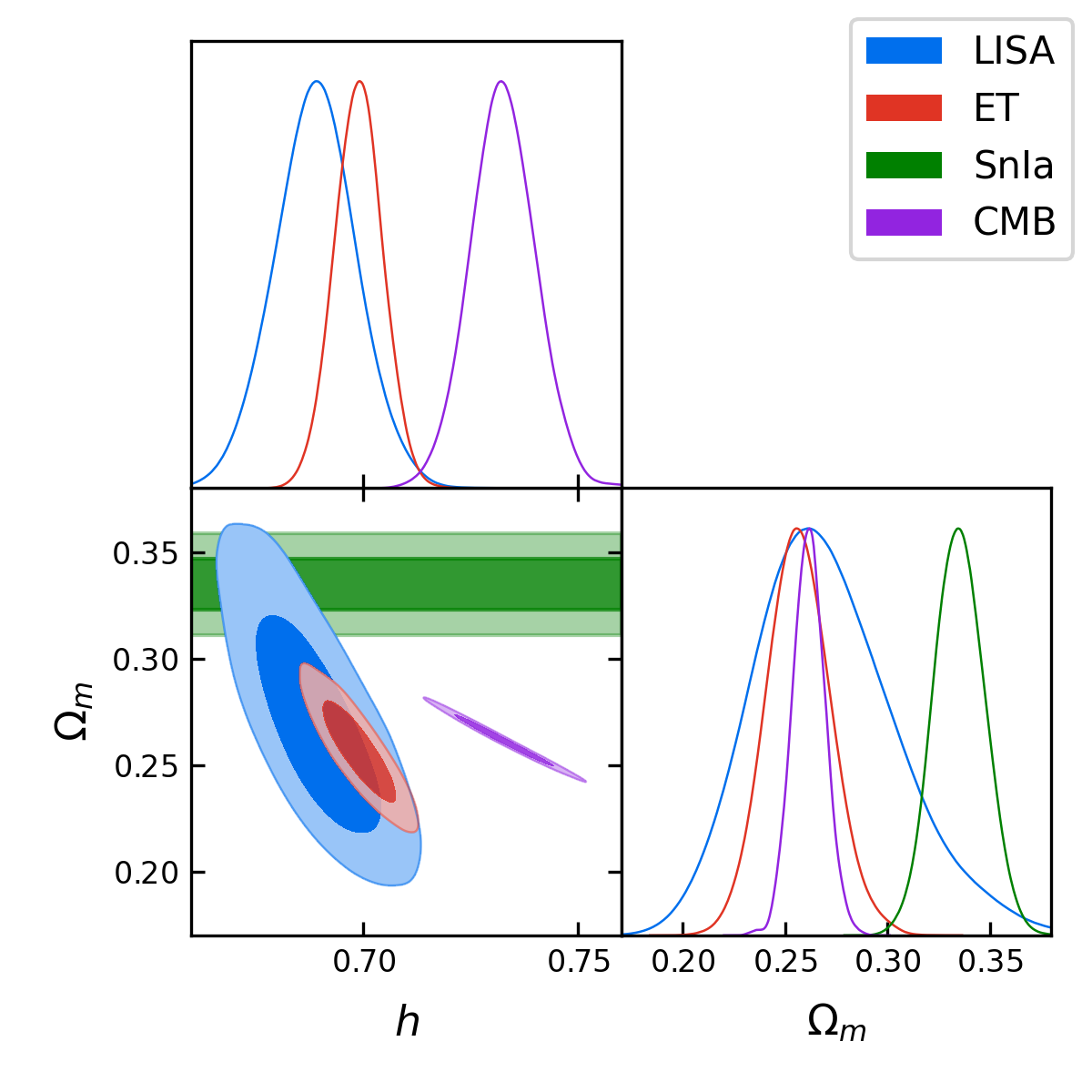}
    \caption{Constraints set by \gls{SnIa}, \gls{CMB} shift parameters and forecast \gls{SS} events for the \gls{ET} and the median case for \gls{LISA}, for the model of $f(Q)$ presented in \cref{eq:model}.}
    \label{fig:fotis-ET-LISA-CMB-SnIa}
\end{figure}

\begin{table}[t]
    \begin{tabular}{ccccc}
        \hline                                                                                             \\[-5pt]
                   & SnIa              & CMB               & LISA                      & ET                \\[4pt]
        \hline\hline                                                                                       \\[-5pt]
        $\Omega_m$ & $0.335 \pm 0.012$ & $0.262 \pm 0.008$ & $0.270^{+0.029}_{-0.039}$ & $0.257 \pm 0.016$ \\[2.5pt]
        $h$        & \rule{1cm}{0.1mm} & $0.732 \pm 0.008$ & $0.689 \pm 0.009$         & $0.699 \pm 0.006$ \\[4pt]
        \hline
    \end{tabular}
    \caption{Best fit and 1$\sigma$ error set by \gls{SnIa}, \gls{CMB} and forecast \gls{SS} events for \gls{LISA} and the \gls{ET}, for the model of $f(Q)$ under study.}
    \label{tab:fotis-ET-LISA-CMB-SnIa}
\end{table}

For the $f(Q)$ model the corner plot is presented in \cref{fig:fotis-ET-LISA-CMB-SnIa} and the best fit values, with the corresponding 1$\sigma$ region, are presented in \cref{tab:fotis-ET-LISA-CMB-SnIa}. We can immediately see that there are tensions in both $\Omega_m$ and $h$ in this model.

The best fit value we obtained for $\Omega_m$ using \gls{SnIa} is consistent with the analysis made in \cite{Anagnostopoulos2021} that also includes baryonic acoustic oscillations and cosmic chronometers. The value $\Omega_m \simeq 0.335$ differs from the one obtained for \gls{LCDM}, $\Omega_m \simeq 0.285$, not surprisingly given that the low redshift \cref{eq:E-model-lowredshift} includes a multiplicative factor $e^{-\lambda}$. 

On the other hand, the \gls{SnIa} value for $\Omega_m$ is in tension with the value we obtain using \gls{CMB}. This happens because the \gls{CMB} takes into account a cumulative effect from high to low redshifts, thus considering both limits \cref{eq:E-model-highredshift} and \cref{eq:E-model-lowredshift}.

We observe that the $\Omega_m$ values of \gls{SnIa} and \gls{SS} are also in tension. It is true that both observables probe the same low redshift regime, however their luminosity distances are different due to the modified gravity factor seen in \cref{eq:dlGW-model}. This extra factor fully explains the tension between the two measurements, and can be seen as a direct test of the existence of a modified gravity term.
We have verified that the tension in $\Omega_m$ between the \gls{SS} events and \gls{SnIa} disappears removing this correction from the luminosity distance of \gls{GW}.
This analysis shows that comparing future \gls{SS} events with standard candles can be decisive in testing modified gravity models using low redshift observables alone.

As in the case of the \gls{SnIa} and \gls{CMB}, we also expect to see a tension arising between the \gls{GW} observables and the \gls{CMB}. This tension, however, depends on the specific choices for the fiducial values used in the \gls{SS} mock data. In this instance, we can see a clear tension between the measurements of $h$, but comparable values for $\Omega_m$. This latter result is just a coincidence occurring for this particular choice of $h$, and is unlikely to be a general feature.

\vspace{-0.3cm}  
\section{Conclusions}
\label{sec:conclusions}

We studied an $f(Q)$ modification of gravity that accounts for the late time accelerated expansion of the Universe without adding a cosmological constant. This modification has the same number of free parameters as the standard \gls{LCDM} model, albeit, the propagation of tensorial perturbations introduces a change in the luminosity distance of \glspl{GW}.

We used a dynamical system analysis to reveal the existence of three fixed points and three disjoint regions in the phase space defined by the sign of the parameter $\lambda$. For $\lambda <0$ we identified a region characterized by trajectories where $E(z) \rightarrow 0$ as the scale factor increases. When $\lambda = 0$, we recover a \gls{CDM} universe with a radiation followed by a matter epoch. When $\lambda > 0$, the stable fixed point corresponds to a regime of eternal inflation, meaning that this case can lead to a viable universe with a late-time accelerated expansion without an explicit cosmological constant.

First we constrained this model using \gls{SnIa} recovering the value of $\Omega_m$ obtained previously in \cite{Anagnostopoulos2021}, where other low redshift observations were also used. This value is, not surprisingly, different from the best fit found for a \gls{LCDM}, given the factor $e^{-\lambda} \Omega_m$ in the modified Friedmann equation \cref{eq:E-model-lowredshift}. 

We made an independent constraint using high redshift measurements from \gls{CMB} and verified that the best fit value for $\Omega_m$ is in tension with the value obtained with \gls{SnIa}. We conclude that this is a manifestation of differing high and low redshift limits of the model, that \gls{CMB} takes into account but \gls{SnIa} does not. 
In order to keep the attractive features of this model at low redshifts, the form of $f(Q)$ needs to be somehow modified at high redshifts to make it consistent with \gls{CMB} data.

As is, the current data disfavors this model already at the background level. This is a common feature in models of modified gravity that depart from \gls{LCDM}.
$f(T)$ and $f(Q)$ models can be shown to be equivalent in \gls{FLRW}, whereas $f(R)$ models, though being formally distinguishable, must in practice have a similar cosmological evolution. In particular, $f(R)$ models have their free parameters strongly constrained to be very close to the \gls{LCDM} limit \cite{1610.07518}, which seems also to be the case for $f(T)$ and $f(Q)$ \cite{2112.15249,Lazkoz2019,Ayuso2020}.

Finally we used mock catalogs of \gls{SS} constructed with \gls{LCDM} as the fiducial model. As expected, the modified luminosity distance for gravitational waves raises a tension with the \gls{SnIa} results implying that in the future, this feature will immediately discriminate this model from \gls{LCDM}.

Although current data (\gls{CMB} and \gls{SnIa}) already disfavor this model, this study  provides a working example of how future \gls{SS} events, measured both by \gls{LISA} and the \gls{ET}, will play a fundamental role when testing models of modified gravity using low redshift data alone.

\vspace{0.4cm}  
\begin{acknowledgments}
The authors are thankful to Fotis Anagnostopoulos for the fruitful discussions.

The authors acknowledge support from Funda\c{c}\~ao para a Ci\^encia e a Tecnologia via the following projects: 
UIDB/04434/2020 \& UIDP/04434/2020,
CERN/FIS-PAR/0037/2019,
PTDC/FIS-OUT/29048/2017,
COMPETE2020: POCI-01-0145-FEDER-028987 \& FCT: PTDC/FIS-AST/28987/2017, PTDC/FIS-AST/0054/2021, EXPL/FIS-AST/1368/2021.
\end{acknowledgments}

\bibliography{bibliography}

\begin{thebibliography}{45}%
\makeatletter
\providecommand \@ifxundefined [1]{%
 \@ifx{#1\undefined}
}%
\providecommand \@ifnum [1]{%
 \ifnum #1\expandafter \@firstoftwo
 \else \expandafter \@secondoftwo
 \fi
}%
\providecommand \@ifx [1]{%
 \ifx #1\expandafter \@firstoftwo
 \else \expandafter \@secondoftwo
 \fi
}%
\providecommand \natexlab [1]{#1}%
\providecommand \enquote  [1]{``#1''}%
\providecommand \bibnamefont  [1]{#1}%
\providecommand \bibfnamefont [1]{#1}%
\providecommand \citenamefont [1]{#1}%
\providecommand \href@noop [0]{\@secondoftwo}%
\providecommand \href [0]{\begingroup \@sanitize@url \@href}%
\providecommand \@href[1]{\@@startlink{#1}\@@href}%
\providecommand \@@href[1]{\endgroup#1\@@endlink}%
\providecommand \@sanitize@url [0]{\catcode `\\12\catcode `\$12\catcode
  `\&12\catcode `\#12\catcode `\^12\catcode `\_12\catcode `\%12\relax}%
\providecommand \@@startlink[1]{}%
\providecommand \@@endlink[0]{}%
\providecommand \url  [0]{\begingroup\@sanitize@url \@url }%
\providecommand \@url [1]{\endgroup\@href {#1}{\urlprefix }}%
\providecommand \urlprefix  [0]{URL }%
\providecommand \Eprint [0]{\href }%
\providecommand \doibase [0]{https://doi.org/}%
\providecommand \selectlanguage [0]{\@gobble}%
\providecommand \bibinfo  [0]{\@secondoftwo}%
\providecommand \bibfield  [0]{\@secondoftwo}%
\providecommand \translation [1]{[#1]}%
\providecommand \BibitemOpen [0]{}%
\providecommand \bibitemStop [0]{}%
\providecommand \bibitemNoStop [0]{.\EOS\space}%
\providecommand \EOS [0]{\spacefactor3000\relax}%
\providecommand \BibitemShut  [1]{\csname bibitem#1\endcsname}%
\let\auto@bib@innerbib\@empty
\bibitem [{\citenamefont {Perivolaropoulos}\ and\ \citenamefont
  {Skara}(2022)}]{LCDM-R2021}%
  \BibitemOpen
  \bibfield  {author} {\bibinfo {author} {\bibfnamefont {L.}~\bibnamefont
  {Perivolaropoulos}}\ and\ \bibinfo {author} {\bibfnamefont {F.}~\bibnamefont
  {Skara}},\ }\bibfield  {title} {\bibinfo {title} {Challenges for
  ${\Lambda}${CDM}: An update},\ }\href
  {https://doi.org/https://doi.org/10.1016/j.newar.2022.101659} {\bibfield
  {journal} {\bibinfo  {journal} {New Astronomy Reviews}\ }\textbf {\bibinfo
  {volume} {95}},\ \bibinfo {pages} {101659} (\bibinfo {year} {2022})},\
  \Eprint {https://arxiv.org/abs/2105.05208} {arXiv:2105.05208 [astro-ph.CO]}
  \BibitemShut {NoStop}%
\bibitem [{\citenamefont {Jim{\'{e}}nez}\ \emph {et~al.}(2019)\citenamefont
  {Jim{\'{e}}nez}, \citenamefont {Heisenberg},\ and\ \citenamefont
  {Koivisto}}]{Jimenez2019a}%
  \BibitemOpen
  \bibfield  {author} {\bibinfo {author} {\bibfnamefont {J.~B.}\ \bibnamefont
  {Jim{\'{e}}nez}}, \bibinfo {author} {\bibfnamefont {L.}~\bibnamefont
  {Heisenberg}},\ and\ \bibinfo {author} {\bibfnamefont {T.}~\bibnamefont
  {Koivisto}},\ }\bibfield  {title} {\bibinfo {title} {The geometrical trinity
  of gravity},\ }\href {https://doi.org/10.3390/universe5070173} {\bibfield
  {journal} {\bibinfo  {journal} {Universe}\ }\textbf {\bibinfo {volume} {5}},\
  \bibinfo {pages} {173} (\bibinfo {year} {2019})},\ \Eprint
  {https://arxiv.org/abs/1903.06830} {arXiv:1903.06830 [hep-th]} \BibitemShut
  {NoStop}%
\bibitem [{\citenamefont {Jimenez}\ \emph {et~al.}(2017)\citenamefont
  {Jimenez}, \citenamefont {Heisenberg},\ and\ \citenamefont
  {Koivisto}}]{Jimenez2017}%
  \BibitemOpen
  \bibfield  {author} {\bibinfo {author} {\bibfnamefont {J.~B.}\ \bibnamefont
  {Jimenez}}, \bibinfo {author} {\bibfnamefont {L.}~\bibnamefont
  {Heisenberg}},\ and\ \bibinfo {author} {\bibfnamefont {T.}~\bibnamefont
  {Koivisto}},\ }\bibfield  {title} {\bibinfo {title} {Coincident {G}eneral
  {R}elativity},\ }\bibfield  {journal} {\bibinfo  {journal} {Phys. Rev. D 98,
  044048 (2018)}\ }\href {https://doi.org/10.1103/PhysRevD.98.044048}
  {10.1103/PhysRevD.98.044048} (\bibinfo {year} {2017}),\ \Eprint
  {https://arxiv.org/abs/1710.03116} {arXiv:1710.03116 [gr-qc]} \BibitemShut
  {NoStop}%
\bibitem [{\citenamefont {Cai}\ and\ \citenamefont {Yang}(2016)}]{Cai2016}%
  \BibitemOpen
  \bibfield  {author} {\bibinfo {author} {\bibfnamefont {R.-G.}\ \bibnamefont
  {Cai}}\ and\ \bibinfo {author} {\bibfnamefont {T.}~\bibnamefont {Yang}},\
  }\bibfield  {title} {\bibinfo {title} {Estimating cosmological parameters by
  the simulated data of gravitational waves from the {E}instein {T}elescope},\
  }\bibfield  {journal} {\bibinfo  {journal} {Phys. Rev. D 95, 044024 (2017)}\
  }\href {https://doi.org/10.1103/PhysRevD.95.044024}
  {10.1103/PhysRevD.95.044024} (\bibinfo {year} {2016}),\ \Eprint
  {https://arxiv.org/abs/1608.08008} {arXiv:1608.08008 [astro-ph.CO]}
  \BibitemShut {NoStop}%
\bibitem [{\citenamefont {Saridakis}\ \emph {et~al.}(2021)\citenamefont
  {Saridakis}, \citenamefont {Lazkoz}, \citenamefont {Salzano}, \citenamefont
  {Moniz}, \citenamefont {Capozziello}, \citenamefont {Jiménez}, \citenamefont
  {Laurentis},\ and\ \citenamefont {Olmo}}]{MGCosmology2021}%
  \BibitemOpen
  \bibinfo {editor} {\bibfnamefont {E.~N.}\ \bibnamefont {Saridakis}}, \bibinfo
  {editor} {\bibfnamefont {R.}~\bibnamefont {Lazkoz}}, \bibinfo {editor}
  {\bibfnamefont {V.}~\bibnamefont {Salzano}}, \bibinfo {editor} {\bibfnamefont
  {P.~V.}\ \bibnamefont {Moniz}}, \bibinfo {editor} {\bibfnamefont
  {S.}~\bibnamefont {Capozziello}}, \bibinfo {editor} {\bibfnamefont {J.~B.}\
  \bibnamefont {Jiménez}}, \bibinfo {editor} {\bibfnamefont {M.~D.}\
  \bibnamefont {Laurentis}},\ and\ \bibinfo {editor} {\bibfnamefont {G.~J.}\
  \bibnamefont {Olmo}},\ eds.,\ \href
  {https://www.ebook.de/de/product/42071306/modified_gravity_and_cosmology.html}
  {\emph {\bibinfo {title} {Modified Gravity and Cosmology}}}\ (\bibinfo
  {publisher} {Springer International Publishing},\ \bibinfo {year}
  {2021})\BibitemShut {NoStop}%
\bibitem [{\citenamefont {Ayuso}\ \emph {et~al.}(2022)\citenamefont {Ayuso},
  \citenamefont {Lazkoz},\ and\ \citenamefont {Mimoso}}]{Ayuso2021}%
  \BibitemOpen
  \bibfield  {author} {\bibinfo {author} {\bibfnamefont {I.}~\bibnamefont
  {Ayuso}}, \bibinfo {author} {\bibfnamefont {R.}~\bibnamefont {Lazkoz}},\ and\
  \bibinfo {author} {\bibfnamefont {J.~P.}\ \bibnamefont {Mimoso}},\ }\bibfield
   {title} {\bibinfo {title} {{DGP} and {DGP}-like cosmologies from $f({Q})$
  actions},\ }\href {https://doi.org/10.1103/PhysRevD.105.083534} {\bibfield
  {journal} {\bibinfo  {journal} {Phys. Rev. D}\ }\textbf {\bibinfo {volume}
  {105}},\ \bibinfo {pages} {083534} (\bibinfo {year} {2022})}\BibitemShut
  {NoStop}%
\bibitem [{\citenamefont {Barros}\ \emph {et~al.}(2020)\citenamefont {Barros},
  \citenamefont {Barreiro}, \citenamefont {Koivisto},\ and\ \citenamefont
  {Nunes}}]{Barros2020}%
  \BibitemOpen
  \bibfield  {author} {\bibinfo {author} {\bibfnamefont {B.~J.}\ \bibnamefont
  {Barros}}, \bibinfo {author} {\bibfnamefont {T.}~\bibnamefont {Barreiro}},
  \bibinfo {author} {\bibfnamefont {T.}~\bibnamefont {Koivisto}},\ and\
  \bibinfo {author} {\bibfnamefont {N.~J.}\ \bibnamefont {Nunes}},\ }\bibfield
  {title} {\bibinfo {title} {{Testing $F({Q})$ gravity with redshift space
  distortions}},\ }\href {https://doi.org/10.1016/j.dark.2020.100616}
  {\bibfield  {journal} {\bibinfo  {journal} {Physics of the Dark Universe}\
  }\textbf {\bibinfo {volume} {30}},\ \bibinfo {pages} {100616} (\bibinfo
  {year} {2020})},\ \Eprint {https://arxiv.org/abs/2004.07867}
  {arXiv:2004.07867 [gr-qc]} \BibitemShut {NoStop}%
\bibitem [{\citenamefont {Belgacem}\ \emph {et~al.}(2017)\citenamefont
  {Belgacem}, \citenamefont {Dirian}, \citenamefont {Foffa},\ and\
  \citenamefont {Maggiore}}]{Belgacem2017a}%
  \BibitemOpen
  \bibfield  {author} {\bibinfo {author} {\bibfnamefont {E.}~\bibnamefont
  {Belgacem}}, \bibinfo {author} {\bibfnamefont {Y.}~\bibnamefont {Dirian}},
  \bibinfo {author} {\bibfnamefont {S.}~\bibnamefont {Foffa}},\ and\ \bibinfo
  {author} {\bibfnamefont {M.}~\bibnamefont {Maggiore}},\ }\bibfield  {title}
  {\bibinfo {title} {The gravitational-wave luminosity distance in modified
  gravity theories},\ }\bibfield  {journal} {\bibinfo  {journal} {Phys. Rev. D
  97, 104066 (2018)}\ }\href {https://doi.org/10.1103/PhysRevD.97.104066}
  {10.1103/PhysRevD.97.104066} (\bibinfo {year} {2017}),\ \Eprint
  {https://arxiv.org/abs/1712.08108} {arXiv:1712.08108 [astro-ph.CO]}
  \BibitemShut {NoStop}%
\bibitem [{\citenamefont {Albuquerque}\ and\ \citenamefont
  {Frusciante}(2022)}]{noemi1}%
  \BibitemOpen
  \bibfield  {author} {\bibinfo {author} {\bibfnamefont {I.~S.}\ \bibnamefont
  {Albuquerque}}\ and\ \bibinfo {author} {\bibfnamefont {N.}~\bibnamefont
  {Frusciante}},\ }\bibfield  {title} {\bibinfo {title} {A designer approach to
  f({Q}) gravity and cosmological implications},\ }\href
  {https://doi.org/https://doi.org/10.1016/j.dark.2022.100980} {\bibfield
  {journal} {\bibinfo  {journal} {Physics of the Dark Universe}\ }\textbf
  {\bibinfo {volume} {35}},\ \bibinfo {pages} {100980} (\bibinfo {year}
  {2022})}\BibitemShut {NoStop}%
\bibitem [{\citenamefont {Atayde}\ and\ \citenamefont
  {Frusciante}(2023)}]{noemi2}%
  \BibitemOpen
  \bibfield  {author} {\bibinfo {author} {\bibfnamefont {L.}~\bibnamefont
  {Atayde}}\ and\ \bibinfo {author} {\bibfnamefont {N.}~\bibnamefont
  {Frusciante}},\ }\bibfield  {title} {\bibinfo {title} {{f(Q) gravity and
  neutrino physics}},\ }\href {https://doi.org/10.1103/PhysRevD.107.124048}
  {\bibfield  {journal} {\bibinfo  {journal} {Phys. Rev. D}\ }\textbf {\bibinfo
  {volume} {107}},\ \bibinfo {pages} {124048} (\bibinfo {year} {2023})},\
  \Eprint {https://arxiv.org/abs/2306.03015} {arXiv:2306.03015 [astro-ph.CO]}
  \BibitemShut {NoStop}%
\bibitem [{\citenamefont {Atayde}\ and\ \citenamefont
  {Frusciante}(2021)}]{noemi3}%
  \BibitemOpen
  \bibfield  {author} {\bibinfo {author} {\bibfnamefont {L.}~\bibnamefont
  {Atayde}}\ and\ \bibinfo {author} {\bibfnamefont {N.}~\bibnamefont
  {Frusciante}},\ }\bibfield  {title} {\bibinfo {title} {Can $f({Q})$ gravity
  challenge $\mathrm{\ensuremath{\Lambda}}\mathrm{CDM}$?},\ }\href
  {https://doi.org/10.1103/PhysRevD.104.064052} {\bibfield  {journal} {\bibinfo
   {journal} {Phys. Rev. D}\ }\textbf {\bibinfo {volume} {104}},\ \bibinfo
  {pages} {064052} (\bibinfo {year} {2021})}\BibitemShut {NoStop}%
\bibitem [{\citenamefont {Frusciante}(2021)}]{noemi4}%
  \BibitemOpen
  \bibfield  {author} {\bibinfo {author} {\bibfnamefont {N.}~\bibnamefont
  {Frusciante}},\ }\bibfield  {title} {\bibinfo {title} {Signatures of $f({Q})$
  gravity in cosmology},\ }\href {https://doi.org/10.1103/PhysRevD.103.044021}
  {\bibfield  {journal} {\bibinfo  {journal} {Phys. Rev. D}\ }\textbf {\bibinfo
  {volume} {103}},\ \bibinfo {pages} {044021} (\bibinfo {year}
  {2021})}\BibitemShut {NoStop}%
\bibitem [{\citenamefont {Capozziello}\ \emph {et~al.}(2022)\citenamefont
  {Capozziello}, \citenamefont {Falco},\ and\ \citenamefont
  {Ferrara}}]{defalco}%
  \BibitemOpen
  \bibfield  {author} {\bibinfo {author} {\bibfnamefont {S.}~\bibnamefont
  {Capozziello}}, \bibinfo {author} {\bibfnamefont {V.~D.}\ \bibnamefont
  {Falco}},\ and\ \bibinfo {author} {\bibfnamefont {C.}~\bibnamefont
  {Ferrara}},\ }\bibfield  {title} {\bibinfo {title} {Comparing equivalent
  gravities: common features and differences},\ }\bibfield  {journal} {\bibinfo
   {journal} {The European Physical Journal C}\ }\textbf {\bibinfo {volume}
  {82}},\ \href {https://doi.org/10.1140/epjc/s10052-022-10823-x}
  {10.1140/epjc/s10052-022-10823-x} (\bibinfo {year} {2022})\BibitemShut
  {NoStop}%
\bibitem [{\citenamefont {Krishnan}\ \emph {et~al.}(2021)\citenamefont
  {Krishnan}, \citenamefont {Colg{\'{a}}in}, \citenamefont {Sheikh-Jabbari},\
  and\ \citenamefont {Yang}}]{colgain}%
  \BibitemOpen
  \bibfield  {author} {\bibinfo {author} {\bibfnamefont {C.}~\bibnamefont
  {Krishnan}}, \bibinfo {author} {\bibfnamefont {E.~{\'{O}}.}\ \bibnamefont
  {Colg{\'{a}}in}}, \bibinfo {author} {\bibfnamefont {M.}~\bibnamefont
  {Sheikh-Jabbari}},\ and\ \bibinfo {author} {\bibfnamefont {T.}~\bibnamefont
  {Yang}},\ }\bibfield  {title} {\bibinfo {title} {Running hubble tension and a
  h0 diagnostic},\ }\bibfield  {journal} {\bibinfo  {journal} {Physical Review
  D}\ }\textbf {\bibinfo {volume} {103}},\ \href
  {https://doi.org/10.1103/physrevd.103.103509} {10.1103/physrevd.103.103509}
  (\bibinfo {year} {2021})\BibitemShut {NoStop}%
\bibitem [{\citenamefont {Khyllep}\ \emph {et~al.}(2021)\citenamefont
  {Khyllep}, \citenamefont {Paliathanasis},\ and\ \citenamefont
  {Dutta}}]{andronikos}%
  \BibitemOpen
  \bibfield  {author} {\bibinfo {author} {\bibfnamefont {W.}~\bibnamefont
  {Khyllep}}, \bibinfo {author} {\bibfnamefont {A.}~\bibnamefont
  {Paliathanasis}},\ and\ \bibinfo {author} {\bibfnamefont {J.}~\bibnamefont
  {Dutta}},\ }\bibfield  {title} {\bibinfo {title} {{Cosmological solutions and
  growth index of matter perturbations in $f(Q)$ gravity}},\ }\href
  {https://doi.org/10.1103/PhysRevD.103.103521} {\bibfield  {journal} {\bibinfo
   {journal} {Phys. Rev. D}\ }\textbf {\bibinfo {volume} {103}},\ \bibinfo
  {pages} {103521} (\bibinfo {year} {2021})},\ \Eprint
  {https://arxiv.org/abs/2103.08372} {arXiv:2103.08372 [gr-qc]} \BibitemShut
  {NoStop}%
\bibitem [{\citenamefont {Ayuso}\ \emph {et~al.}(2021)\citenamefont {Ayuso},
  \citenamefont {Lazkoz},\ and\ \citenamefont {Salzano}}]{Ayuso2020}%
  \BibitemOpen
  \bibfield  {author} {\bibinfo {author} {\bibfnamefont {I.}~\bibnamefont
  {Ayuso}}, \bibinfo {author} {\bibfnamefont {R.}~\bibnamefont {Lazkoz}},\ and\
  \bibinfo {author} {\bibfnamefont {V.}~\bibnamefont {Salzano}},\ }\bibfield
  {title} {\bibinfo {title} {{Observational constraints on cosmological
  solutions of $f({Q})$ theories}},\ }\href
  {https://doi.org/10.1103/PhysRevD.103.063505} {\bibfield  {journal} {\bibinfo
   {journal} {Phys. Rev. D}\ }\textbf {\bibinfo {volume} {103}},\ \bibinfo
  {pages} {063505} (\bibinfo {year} {2021})},\ \Eprint
  {https://arxiv.org/abs/2012.00046} {2012.00046 [astro-ph.CO]} \BibitemShut
  {NoStop}%
\bibitem [{\citenamefont {Collaboration}\ and\ \citenamefont
  {Collaboration}(2017)}]{GW170817}%
  \BibitemOpen
  \bibfield  {author} {\bibinfo {author} {\bibfnamefont {T.~L.~S.}\
  \bibnamefont {Collaboration}}\ and\ \bibinfo {author} {\bibfnamefont {T.~V.}\
  \bibnamefont {Collaboration}},\ }\bibfield  {title} {\bibinfo {title}
  {Gw170817: Observation of gravitational waves from a binary neutron star
  inspiral},\ }\bibfield  {journal} {\bibinfo  {journal} {Phys. Rev. Lett. 119
  161101 (2017)}\ }\href {https://doi.org/10.1103/PhysRevLett.119.161101}
  {10.1103/PhysRevLett.119.161101} (\bibinfo {year} {2017}),\ \Eprint
  {https://arxiv.org/abs/1710.05832} {arXiv:1710.05832 [gr-qc]} \BibitemShut
  {NoStop}%
\bibitem [{\citenamefont {Collaboration}\ and\ \citenamefont {the
  Virgo~Collaboration}(2020)}]{GW190521}%
  \BibitemOpen
  \bibfield  {author} {\bibinfo {author} {\bibfnamefont {T.~L.~S.}\
  \bibnamefont {Collaboration}}\ and\ \bibinfo {author} {\bibnamefont {the
  Virgo~Collaboration}},\ }\bibfield  {title} {\bibinfo {title} {Properties and
  astrophysical implications of the 150 msun binary black hole merger
  {GW}190521},\ }\bibfield  {journal} {\bibinfo  {journal} {Astrophys. J. Lett.
  900, L13 (2020)}\ }\href {https://doi.org/10.3847/2041-8213/aba493}
  {10.3847/2041-8213/aba493} (\bibinfo {year} {2020}),\ \Eprint
  {https://arxiv.org/abs/2009.01190} {arXiv:2009.01190 [astro-ph.HE]}
  \BibitemShut {NoStop}%
\bibitem [{\citenamefont {Graham}\ \emph {et~al.}(2020)\citenamefont {Graham},
  \citenamefont {Ford}, \citenamefont {McKernan}, \citenamefont {Ross},
  \citenamefont {Stern}, \citenamefont {Burdge}, \citenamefont {Coughlin},
  \citenamefont {Djorgovski}, \citenamefont {Drake}, \citenamefont {Duev},
  \citenamefont {Kasliwal}, \citenamefont {Mahabal}, \citenamefont {van
  Velzen}, \citenamefont {Belicki}, \citenamefont {Bellm}, \citenamefont
  {Burruss}, \citenamefont {Cenko}, \citenamefont {Cunningham}, \citenamefont
  {Helou}, \citenamefont {Kulkarni}, \citenamefont {Masci}, \citenamefont
  {Prince}, \citenamefont {Reiley}, \citenamefont {Rodriguez}, \citenamefont
  {Rusholme}, \citenamefont {Smith},\ and\ \citenamefont
  {Soumagnac}}]{GW190521-EM}%
  \BibitemOpen
  \bibfield  {author} {\bibinfo {author} {\bibfnamefont {M.~J.}\ \bibnamefont
  {Graham}}, \bibinfo {author} {\bibfnamefont {K.~E.~S.}\ \bibnamefont {Ford}},
  \bibinfo {author} {\bibfnamefont {B.}~\bibnamefont {McKernan}}, \bibinfo
  {author} {\bibfnamefont {N.~P.}\ \bibnamefont {Ross}}, \bibinfo {author}
  {\bibfnamefont {D.}~\bibnamefont {Stern}}, \bibinfo {author} {\bibfnamefont
  {K.}~\bibnamefont {Burdge}}, \bibinfo {author} {\bibfnamefont
  {M.}~\bibnamefont {Coughlin}}, \bibinfo {author} {\bibfnamefont {S.~G.}\
  \bibnamefont {Djorgovski}}, \bibinfo {author} {\bibfnamefont {A.~J.}\
  \bibnamefont {Drake}}, \bibinfo {author} {\bibfnamefont {D.}~\bibnamefont
  {Duev}}, \bibinfo {author} {\bibfnamefont {M.}~\bibnamefont {Kasliwal}},
  \bibinfo {author} {\bibfnamefont {A.~A.}\ \bibnamefont {Mahabal}}, \bibinfo
  {author} {\bibfnamefont {S.}~\bibnamefont {van Velzen}}, \bibinfo {author}
  {\bibfnamefont {J.}~\bibnamefont {Belicki}}, \bibinfo {author} {\bibfnamefont
  {E.~C.}\ \bibnamefont {Bellm}}, \bibinfo {author} {\bibfnamefont
  {R.}~\bibnamefont {Burruss}}, \bibinfo {author} {\bibfnamefont {S.~B.}\
  \bibnamefont {Cenko}}, \bibinfo {author} {\bibfnamefont {V.}~\bibnamefont
  {Cunningham}}, \bibinfo {author} {\bibfnamefont {G.}~\bibnamefont {Helou}},
  \bibinfo {author} {\bibfnamefont {S.~R.}\ \bibnamefont {Kulkarni}}, \bibinfo
  {author} {\bibfnamefont {F.~J.}\ \bibnamefont {Masci}}, \bibinfo {author}
  {\bibfnamefont {T.}~\bibnamefont {Prince}}, \bibinfo {author} {\bibfnamefont
  {D.}~\bibnamefont {Reiley}}, \bibinfo {author} {\bibfnamefont
  {H.}~\bibnamefont {Rodriguez}}, \bibinfo {author} {\bibfnamefont
  {B.}~\bibnamefont {Rusholme}}, \bibinfo {author} {\bibfnamefont {R.~M.}\
  \bibnamefont {Smith}},\ and\ \bibinfo {author} {\bibfnamefont {M.~T.}\
  \bibnamefont {Soumagnac}},\ }\bibfield  {title} {\bibinfo {title} {Candidate
  electromagnetic counterpart to the binary black hole merger gravitational
  wave event {S}190521g},\ }\bibfield  {journal} {\bibinfo  {journal} {Phys.
  Rev. Lett. 124, 251102}\ }\href
  {https://doi.org/10.1103/PhysRevLett.124.251102}
  {10.1103/PhysRevLett.124.251102} (\bibinfo {year} {2020}),\ \Eprint
  {https://arxiv.org/abs/2006.14122} {arXiv:2006.14122 [astro-ph.HE]}
  \BibitemShut {NoStop}%
\bibitem [{\citenamefont {Amaro-Seoane}\ \emph {et~al.}(2017)\citenamefont
  {Amaro-Seoane}, \citenamefont {Audley}, \citenamefont {Babak}, \citenamefont
  {Baker}, \citenamefont {Barausse}, \citenamefont {Bender}, \citenamefont
  {Berti}, \citenamefont {Binetruy}, \citenamefont {Born}, \citenamefont
  {Bortoluzzi}, \citenamefont {Camp}, \citenamefont {Caprini}, \citenamefont
  {Cardoso}, \citenamefont {Colpi}, \citenamefont {Conklin}, \citenamefont
  {Cornish}, \citenamefont {Cutler}, \citenamefont {Danzmann}, \citenamefont
  {Dolesi}, \citenamefont {Ferraioli}, \citenamefont {Ferroni}, \citenamefont
  {Fitzsimons}, \citenamefont {Gair}, \citenamefont {Bote}, \citenamefont
  {Giardini}, \citenamefont {Gibert}, \citenamefont {Grimani}, \citenamefont
  {Halloin}, \citenamefont {Heinzel}, \citenamefont {Hertog}, \citenamefont
  {Hewitson}, \citenamefont {Holley-Bockelmann}, \citenamefont {Hollington},
  \citenamefont {Hueller}, \citenamefont {Inchauspe}, \citenamefont {Jetzer},
  \citenamefont {Karnesis}, \citenamefont {Killow}, \citenamefont {Klein},
  \citenamefont {Klipstein}, \citenamefont {Korsakova}, \citenamefont {Larson},
  \citenamefont {Livas}, \citenamefont {Lloro}, \citenamefont {Man},
  \citenamefont {Mance}, \citenamefont {Martino}, \citenamefont {Mateos},
  \citenamefont {McKenzie}, \citenamefont {McWilliams}, \citenamefont {Miller},
  \citenamefont {Mueller}, \citenamefont {Nardini}, \citenamefont {Nelemans},
  \citenamefont {Nofrarias}, \citenamefont {Petiteau}, \citenamefont {Pivato},
  \citenamefont {Plagnol}, \citenamefont {Porter}, \citenamefont {Reiche},
  \citenamefont {Robertson}, \citenamefont {Robertson}, \citenamefont {Rossi},
  \citenamefont {Russano}, \citenamefont {Schutz}, \citenamefont {Sesana},
  \citenamefont {Shoemaker}, \citenamefont {Slutsky}, \citenamefont {Sopuerta},
  \citenamefont {Sumner}, \citenamefont {Tamanini}, \citenamefont {Thorpe},
  \citenamefont {Troebs}, \citenamefont {Vallisneri}, \citenamefont {Vecchio},
  \citenamefont {Vetrugno}, \citenamefont {Vitale}, \citenamefont {Volonteri},
  \citenamefont {Wanner}, \citenamefont {Ward}, \citenamefont {Wass},
  \citenamefont {Weber}, \citenamefont {Ziemer},\ and\ \citenamefont
  {Zweifel}}]{LISA-proposal}%
  \BibitemOpen
  \bibfield  {author} {\bibinfo {author} {\bibfnamefont {P.}~\bibnamefont
  {Amaro-Seoane}}, \bibinfo {author} {\bibfnamefont {H.}~\bibnamefont
  {Audley}}, \bibinfo {author} {\bibfnamefont {S.}~\bibnamefont {Babak}},
  \bibinfo {author} {\bibfnamefont {J.}~\bibnamefont {Baker}}, \bibinfo
  {author} {\bibfnamefont {E.}~\bibnamefont {Barausse}}, \bibinfo {author}
  {\bibfnamefont {P.}~\bibnamefont {Bender}}, \bibinfo {author} {\bibfnamefont
  {E.}~\bibnamefont {Berti}}, \bibinfo {author} {\bibfnamefont
  {P.}~\bibnamefont {Binetruy}}, \bibinfo {author} {\bibfnamefont
  {M.}~\bibnamefont {Born}}, \bibinfo {author} {\bibfnamefont {D.}~\bibnamefont
  {Bortoluzzi}}, \bibinfo {author} {\bibfnamefont {J.}~\bibnamefont {Camp}},
  \bibinfo {author} {\bibfnamefont {C.}~\bibnamefont {Caprini}}, \bibinfo
  {author} {\bibfnamefont {V.}~\bibnamefont {Cardoso}}, \bibinfo {author}
  {\bibfnamefont {M.}~\bibnamefont {Colpi}}, \bibinfo {author} {\bibfnamefont
  {J.}~\bibnamefont {Conklin}}, \bibinfo {author} {\bibfnamefont
  {N.}~\bibnamefont {Cornish}}, \bibinfo {author} {\bibfnamefont
  {C.}~\bibnamefont {Cutler}}, \bibinfo {author} {\bibfnamefont
  {K.}~\bibnamefont {Danzmann}}, \bibinfo {author} {\bibfnamefont
  {R.}~\bibnamefont {Dolesi}}, \bibinfo {author} {\bibfnamefont
  {L.}~\bibnamefont {Ferraioli}}, \bibinfo {author} {\bibfnamefont
  {V.}~\bibnamefont {Ferroni}}, \bibinfo {author} {\bibfnamefont
  {E.}~\bibnamefont {Fitzsimons}}, \bibinfo {author} {\bibfnamefont
  {J.}~\bibnamefont {Gair}}, \bibinfo {author} {\bibfnamefont {L.~G.}\
  \bibnamefont {Bote}}, \bibinfo {author} {\bibfnamefont {D.}~\bibnamefont
  {Giardini}}, \bibinfo {author} {\bibfnamefont {F.}~\bibnamefont {Gibert}},
  \bibinfo {author} {\bibfnamefont {C.}~\bibnamefont {Grimani}}, \bibinfo
  {author} {\bibfnamefont {H.}~\bibnamefont {Halloin}}, \bibinfo {author}
  {\bibfnamefont {G.}~\bibnamefont {Heinzel}}, \bibinfo {author} {\bibfnamefont
  {T.}~\bibnamefont {Hertog}}, \bibinfo {author} {\bibfnamefont
  {M.}~\bibnamefont {Hewitson}}, \bibinfo {author} {\bibfnamefont
  {K.}~\bibnamefont {Holley-Bockelmann}}, \bibinfo {author} {\bibfnamefont
  {D.}~\bibnamefont {Hollington}}, \bibinfo {author} {\bibfnamefont
  {M.}~\bibnamefont {Hueller}}, \bibinfo {author} {\bibfnamefont
  {H.}~\bibnamefont {Inchauspe}}, \bibinfo {author} {\bibfnamefont
  {P.}~\bibnamefont {Jetzer}}, \bibinfo {author} {\bibfnamefont
  {N.}~\bibnamefont {Karnesis}}, \bibinfo {author} {\bibfnamefont
  {C.}~\bibnamefont {Killow}}, \bibinfo {author} {\bibfnamefont
  {A.}~\bibnamefont {Klein}}, \bibinfo {author} {\bibfnamefont
  {B.}~\bibnamefont {Klipstein}}, \bibinfo {author} {\bibfnamefont
  {N.}~\bibnamefont {Korsakova}}, \bibinfo {author} {\bibfnamefont {S.~L.}\
  \bibnamefont {Larson}}, \bibinfo {author} {\bibfnamefont {J.}~\bibnamefont
  {Livas}}, \bibinfo {author} {\bibfnamefont {I.}~\bibnamefont {Lloro}},
  \bibinfo {author} {\bibfnamefont {N.}~\bibnamefont {Man}}, \bibinfo {author}
  {\bibfnamefont {D.}~\bibnamefont {Mance}}, \bibinfo {author} {\bibfnamefont
  {J.}~\bibnamefont {Martino}}, \bibinfo {author} {\bibfnamefont
  {I.}~\bibnamefont {Mateos}}, \bibinfo {author} {\bibfnamefont
  {K.}~\bibnamefont {McKenzie}}, \bibinfo {author} {\bibfnamefont {S.~T.}\
  \bibnamefont {McWilliams}}, \bibinfo {author} {\bibfnamefont
  {C.}~\bibnamefont {Miller}}, \bibinfo {author} {\bibfnamefont
  {G.}~\bibnamefont {Mueller}}, \bibinfo {author} {\bibfnamefont
  {G.}~\bibnamefont {Nardini}}, \bibinfo {author} {\bibfnamefont
  {G.}~\bibnamefont {Nelemans}}, \bibinfo {author} {\bibfnamefont
  {M.}~\bibnamefont {Nofrarias}}, \bibinfo {author} {\bibfnamefont
  {A.}~\bibnamefont {Petiteau}}, \bibinfo {author} {\bibfnamefont
  {P.}~\bibnamefont {Pivato}}, \bibinfo {author} {\bibfnamefont
  {E.}~\bibnamefont {Plagnol}}, \bibinfo {author} {\bibfnamefont
  {E.}~\bibnamefont {Porter}}, \bibinfo {author} {\bibfnamefont
  {J.}~\bibnamefont {Reiche}}, \bibinfo {author} {\bibfnamefont
  {D.}~\bibnamefont {Robertson}}, \bibinfo {author} {\bibfnamefont
  {N.}~\bibnamefont {Robertson}}, \bibinfo {author} {\bibfnamefont
  {E.}~\bibnamefont {Rossi}}, \bibinfo {author} {\bibfnamefont
  {G.}~\bibnamefont {Russano}}, \bibinfo {author} {\bibfnamefont
  {B.}~\bibnamefont {Schutz}}, \bibinfo {author} {\bibfnamefont
  {A.}~\bibnamefont {Sesana}}, \bibinfo {author} {\bibfnamefont
  {D.}~\bibnamefont {Shoemaker}}, \bibinfo {author} {\bibfnamefont
  {J.}~\bibnamefont {Slutsky}}, \bibinfo {author} {\bibfnamefont {C.~F.}\
  \bibnamefont {Sopuerta}}, \bibinfo {author} {\bibfnamefont {T.}~\bibnamefont
  {Sumner}}, \bibinfo {author} {\bibfnamefont {N.}~\bibnamefont {Tamanini}},
  \bibinfo {author} {\bibfnamefont {I.}~\bibnamefont {Thorpe}}, \bibinfo
  {author} {\bibfnamefont {M.}~\bibnamefont {Troebs}}, \bibinfo {author}
  {\bibfnamefont {M.}~\bibnamefont {Vallisneri}}, \bibinfo {author}
  {\bibfnamefont {A.}~\bibnamefont {Vecchio}}, \bibinfo {author} {\bibfnamefont
  {D.}~\bibnamefont {Vetrugno}}, \bibinfo {author} {\bibfnamefont
  {S.}~\bibnamefont {Vitale}}, \bibinfo {author} {\bibfnamefont
  {M.}~\bibnamefont {Volonteri}}, \bibinfo {author} {\bibfnamefont
  {G.}~\bibnamefont {Wanner}}, \bibinfo {author} {\bibfnamefont
  {H.}~\bibnamefont {Ward}}, \bibinfo {author} {\bibfnamefont {P.}~\bibnamefont
  {Wass}}, \bibinfo {author} {\bibfnamefont {W.}~\bibnamefont {Weber}},
  \bibinfo {author} {\bibfnamefont {J.}~\bibnamefont {Ziemer}},\ and\ \bibinfo
  {author} {\bibfnamefont {P.}~\bibnamefont {Zweifel}},\ }\bibfield  {title}
  {\bibinfo {title} {Laser {I}nterferometer {S}pace {A}ntenna},\ }\href@noop {}
  {\  (\bibinfo {year} {2017})},\ \Eprint {https://arxiv.org/abs/1702.00786}
  {arXiv:1702.00786 [astro-ph.IM]} \BibitemShut {NoStop}%
\bibitem [{\citenamefont {Auclair}\ \emph {et~al.}(2023)\citenamefont
  {Auclair}, \citenamefont {Bacon}, \citenamefont {Baker}, \citenamefont
  {Barreiro}, \citenamefont {Bartolo}, \citenamefont {Belgacem}, \citenamefont
  {Bellomo}, \citenamefont {Ben-Dayan}, \citenamefont {Bertacca}, \citenamefont
  {Besancon}, \citenamefont {Blanco-Pillado}, \citenamefont {Blas},
  \citenamefont {Boileau}, \citenamefont {Calcagni}, \citenamefont {Caldwell},
  \citenamefont {Caprini}, \citenamefont {Carbone}, \citenamefont {Chang},
  \citenamefont {Chen}, \citenamefont {Christensen}, \citenamefont {Clesse},
  \citenamefont {Comelli}, \citenamefont {Congedo}, \citenamefont {Contaldi},
  \citenamefont {Crisostomi}, \citenamefont {Croon}, \citenamefont {Cui},
  \citenamefont {Cusin}, \citenamefont {Cutting}, \citenamefont {Dalang},
  \citenamefont {De~Luca}, \citenamefont {Pozzo}, \citenamefont {Desjacques},
  \citenamefont {Dimastrogiovanni}, \citenamefont {Dorsch}, \citenamefont
  {Ezquiaga}, \citenamefont {Fasiello}, \citenamefont {Figueroa}, \citenamefont
  {Flauger}, \citenamefont {Franciolini}, \citenamefont {Frusciante},
  \citenamefont {Fumagalli}, \citenamefont {Garc{\'i}a-Bellido}, \citenamefont
  {Gould}, \citenamefont {Holz}, \citenamefont {Iacconi}, \citenamefont {Jain},
  \citenamefont {Jenkins}, \citenamefont {Jinno}, \citenamefont {Joana},
  \citenamefont {Karnesis}, \citenamefont {Konstandin}, \citenamefont {Koyama},
  \citenamefont {Kozaczuk}, \citenamefont {Kuroyanagi}, \citenamefont {Laghi},
  \citenamefont {Lewicki}, \citenamefont {Lombriser}, \citenamefont {Madge},
  \citenamefont {Maggiore}, \citenamefont {Malhotra}, \citenamefont
  {Mancarella}, \citenamefont {Mandic}, \citenamefont {Mangiagli},
  \citenamefont {Matarrese}, \citenamefont {Mazumdar}, \citenamefont
  {Mukherjee}, \citenamefont {Musco}, \citenamefont {Nardini}, \citenamefont
  {No}, \citenamefont {Papanikolaou}, \citenamefont {Peloso}, \citenamefont
  {Pieroni}, \citenamefont {Pilo}, \citenamefont {Raccanelli}, \citenamefont
  {Renaux-Petel}, \citenamefont {Renzini}, \citenamefont {Ricciardone},
  \citenamefont {Riotto}, \citenamefont {Romano}, \citenamefont {Rollo},
  \citenamefont {Pol}, \citenamefont {Morales}, \citenamefont {Sakellariadou},
  \citenamefont {Saltas}, \citenamefont {Scalisi}, \citenamefont {Schmitz},
  \citenamefont {Schwaller}, \citenamefont {Sergijenko}, \citenamefont
  {Servant}, \citenamefont {Simakachorn}, \citenamefont {Sorbo}, \citenamefont
  {Sousa}, \citenamefont {Speri}, \citenamefont {Steer}, \citenamefont
  {Tamanini}, \citenamefont {Tasinato}, \citenamefont {Torrado}, \citenamefont
  {Unal}, \citenamefont {Vennin}, \citenamefont {Vernieri}, \citenamefont
  {Vernizzi}, \citenamefont {Volonteri}, \citenamefont {Wachter}, \citenamefont
  {Wands}, \citenamefont {Witkowski}, \citenamefont {Zumalac{\'a}rregui},
  \citenamefont {Annis}, \citenamefont {Ares}, \citenamefont {Avelino},
  \citenamefont {Avgoustidis}, \citenamefont {Barausse}, \citenamefont
  {Bonilla}, \citenamefont {Bonvin}, \citenamefont {Bosso}, \citenamefont
  {Calabrese}, \citenamefont {{\c{C}}al{\i}{\c{s}}kan}, \citenamefont
  {Cembranos}, \citenamefont {Chala}, \citenamefont {Chernoff}, \citenamefont
  {Clough}, \citenamefont {Criswell}, \citenamefont {Das}, \citenamefont
  {Silva}, \citenamefont {Dayal}, \citenamefont {Domcke}, \citenamefont
  {Durrer}, \citenamefont {Easther}, \citenamefont {Escoffier}, \citenamefont
  {Ferrans}, \citenamefont {Fryer}, \citenamefont {Gair}, \citenamefont
  {Gordon}, \citenamefont {Hendry}, \citenamefont {Hindmarsh}, \citenamefont
  {Hooper}, \citenamefont {Kajfasz}, \citenamefont {Kopp}, \citenamefont
  {Koushiappas}, \citenamefont {Kumar}, \citenamefont {Kunz}, \citenamefont
  {Lagos}, \citenamefont {Lilley}, \citenamefont {Lizarraga}, \citenamefont
  {Lobo}, \citenamefont {Maleknejad}, \citenamefont {Martins}, \citenamefont
  {Meerburg}, \citenamefont {Meyer}, \citenamefont {Mimoso}, \citenamefont
  {Nesseris}, \citenamefont {Nunes}, \citenamefont {Oikonomou}, \citenamefont
  {Orlando}, \citenamefont {{\"O}zsoy}, \citenamefont {Pacucci}, \citenamefont
  {Palmese}, \citenamefont {Petiteau}, \citenamefont {Pinol}, \citenamefont
  {Zwart}, \citenamefont {Pratten}, \citenamefont {Prokopec}, \citenamefont
  {Quenby}, \citenamefont {Rastgoo}, \citenamefont {Roest}, \citenamefont
  {Rummukainen}, \citenamefont {Schimd}, \citenamefont {Secroun}, \citenamefont
  {Sesana}, \citenamefont {Sopuerta}, \citenamefont {Tereno}, \citenamefont
  {Tolley}, \citenamefont {Urrestilla}, \citenamefont {Vagenas}, \citenamefont
  {van~de Vis}, \citenamefont {van~de Weygaert}, \citenamefont {Wardell},
  \citenamefont {Weir}, \citenamefont {White}, \citenamefont
  {{\'{S}}wie{\.{z}}ewska}, \citenamefont {Zhdanov},\ and\ \citenamefont
  {Group}}]{Auclair2022}%
  \BibitemOpen
  \bibfield  {author} {\bibinfo {author} {\bibfnamefont {P.}~\bibnamefont
  {Auclair}}, \bibinfo {author} {\bibfnamefont {D.}~\bibnamefont {Bacon}},
  \bibinfo {author} {\bibfnamefont {T.}~\bibnamefont {Baker}}, \bibinfo
  {author} {\bibfnamefont {T.}~\bibnamefont {Barreiro}}, \bibinfo {author}
  {\bibfnamefont {N.}~\bibnamefont {Bartolo}}, \bibinfo {author} {\bibfnamefont
  {E.}~\bibnamefont {Belgacem}}, \bibinfo {author} {\bibfnamefont
  {N.}~\bibnamefont {Bellomo}}, \bibinfo {author} {\bibfnamefont
  {I.}~\bibnamefont {Ben-Dayan}}, \bibinfo {author} {\bibfnamefont
  {D.}~\bibnamefont {Bertacca}}, \bibinfo {author} {\bibfnamefont
  {M.}~\bibnamefont {Besancon}}, \bibinfo {author} {\bibfnamefont {J.~J.}\
  \bibnamefont {Blanco-Pillado}}, \bibinfo {author} {\bibfnamefont
  {D.}~\bibnamefont {Blas}}, \bibinfo {author} {\bibfnamefont {G.}~\bibnamefont
  {Boileau}}, \bibinfo {author} {\bibfnamefont {G.}~\bibnamefont {Calcagni}},
  \bibinfo {author} {\bibfnamefont {R.}~\bibnamefont {Caldwell}}, \bibinfo
  {author} {\bibfnamefont {C.}~\bibnamefont {Caprini}}, \bibinfo {author}
  {\bibfnamefont {C.}~\bibnamefont {Carbone}}, \bibinfo {author} {\bibfnamefont
  {C.-F.}\ \bibnamefont {Chang}}, \bibinfo {author} {\bibfnamefont {H.-Y.}\
  \bibnamefont {Chen}}, \bibinfo {author} {\bibfnamefont {N.}~\bibnamefont
  {Christensen}}, \bibinfo {author} {\bibfnamefont {S.}~\bibnamefont {Clesse}},
  \bibinfo {author} {\bibfnamefont {D.}~\bibnamefont {Comelli}}, \bibinfo
  {author} {\bibfnamefont {G.}~\bibnamefont {Congedo}}, \bibinfo {author}
  {\bibfnamefont {C.}~\bibnamefont {Contaldi}}, \bibinfo {author}
  {\bibfnamefont {M.}~\bibnamefont {Crisostomi}}, \bibinfo {author}
  {\bibfnamefont {D.}~\bibnamefont {Croon}}, \bibinfo {author} {\bibfnamefont
  {Y.}~\bibnamefont {Cui}}, \bibinfo {author} {\bibfnamefont {G.}~\bibnamefont
  {Cusin}}, \bibinfo {author} {\bibfnamefont {D.}~\bibnamefont {Cutting}},
  \bibinfo {author} {\bibfnamefont {C.}~\bibnamefont {Dalang}}, \bibinfo
  {author} {\bibfnamefont {V.}~\bibnamefont {De~Luca}}, \bibinfo {author}
  {\bibfnamefont {W.~D.}\ \bibnamefont {Pozzo}}, \bibinfo {author}
  {\bibfnamefont {V.}~\bibnamefont {Desjacques}}, \bibinfo {author}
  {\bibfnamefont {E.}~\bibnamefont {Dimastrogiovanni}}, \bibinfo {author}
  {\bibfnamefont {G.~C.}\ \bibnamefont {Dorsch}}, \bibinfo {author}
  {\bibfnamefont {J.~M.}\ \bibnamefont {Ezquiaga}}, \bibinfo {author}
  {\bibfnamefont {M.}~\bibnamefont {Fasiello}}, \bibinfo {author}
  {\bibfnamefont {D.~G.}\ \bibnamefont {Figueroa}}, \bibinfo {author}
  {\bibfnamefont {R.}~\bibnamefont {Flauger}}, \bibinfo {author} {\bibfnamefont
  {G.}~\bibnamefont {Franciolini}}, \bibinfo {author} {\bibfnamefont
  {N.}~\bibnamefont {Frusciante}}, \bibinfo {author} {\bibfnamefont
  {J.}~\bibnamefont {Fumagalli}}, \bibinfo {author} {\bibfnamefont
  {J.}~\bibnamefont {Garc{\'i}a-Bellido}}, \bibinfo {author} {\bibfnamefont
  {O.}~\bibnamefont {Gould}}, \bibinfo {author} {\bibfnamefont
  {D.}~\bibnamefont {Holz}}, \bibinfo {author} {\bibfnamefont {L.}~\bibnamefont
  {Iacconi}}, \bibinfo {author} {\bibfnamefont {R.~K.}\ \bibnamefont {Jain}},
  \bibinfo {author} {\bibfnamefont {A.~C.}\ \bibnamefont {Jenkins}}, \bibinfo
  {author} {\bibfnamefont {R.}~\bibnamefont {Jinno}}, \bibinfo {author}
  {\bibfnamefont {C.}~\bibnamefont {Joana}}, \bibinfo {author} {\bibfnamefont
  {N.}~\bibnamefont {Karnesis}}, \bibinfo {author} {\bibfnamefont
  {T.}~\bibnamefont {Konstandin}}, \bibinfo {author} {\bibfnamefont
  {K.}~\bibnamefont {Koyama}}, \bibinfo {author} {\bibfnamefont
  {J.}~\bibnamefont {Kozaczuk}}, \bibinfo {author} {\bibfnamefont
  {S.}~\bibnamefont {Kuroyanagi}}, \bibinfo {author} {\bibfnamefont
  {D.}~\bibnamefont {Laghi}}, \bibinfo {author} {\bibfnamefont
  {M.}~\bibnamefont {Lewicki}}, \bibinfo {author} {\bibfnamefont
  {L.}~\bibnamefont {Lombriser}}, \bibinfo {author} {\bibfnamefont
  {E.}~\bibnamefont {Madge}}, \bibinfo {author} {\bibfnamefont
  {M.}~\bibnamefont {Maggiore}}, \bibinfo {author} {\bibfnamefont
  {A.}~\bibnamefont {Malhotra}}, \bibinfo {author} {\bibfnamefont
  {M.}~\bibnamefont {Mancarella}}, \bibinfo {author} {\bibfnamefont
  {V.}~\bibnamefont {Mandic}}, \bibinfo {author} {\bibfnamefont
  {A.}~\bibnamefont {Mangiagli}}, \bibinfo {author} {\bibfnamefont
  {S.}~\bibnamefont {Matarrese}}, \bibinfo {author} {\bibfnamefont
  {A.}~\bibnamefont {Mazumdar}}, \bibinfo {author} {\bibfnamefont
  {S.}~\bibnamefont {Mukherjee}}, \bibinfo {author} {\bibfnamefont
  {I.}~\bibnamefont {Musco}}, \bibinfo {author} {\bibfnamefont
  {G.}~\bibnamefont {Nardini}}, \bibinfo {author} {\bibfnamefont {J.~M.}\
  \bibnamefont {No}}, \bibinfo {author} {\bibfnamefont {T.}~\bibnamefont
  {Papanikolaou}}, \bibinfo {author} {\bibfnamefont {M.}~\bibnamefont
  {Peloso}}, \bibinfo {author} {\bibfnamefont {M.}~\bibnamefont {Pieroni}},
  \bibinfo {author} {\bibfnamefont {L.}~\bibnamefont {Pilo}}, \bibinfo {author}
  {\bibfnamefont {A.}~\bibnamefont {Raccanelli}}, \bibinfo {author}
  {\bibfnamefont {S.}~\bibnamefont {Renaux-Petel}}, \bibinfo {author}
  {\bibfnamefont {A.~I.}\ \bibnamefont {Renzini}}, \bibinfo {author}
  {\bibfnamefont {A.}~\bibnamefont {Ricciardone}}, \bibinfo {author}
  {\bibfnamefont {A.}~\bibnamefont {Riotto}}, \bibinfo {author} {\bibfnamefont
  {J.~D.}\ \bibnamefont {Romano}}, \bibinfo {author} {\bibfnamefont
  {R.}~\bibnamefont {Rollo}}, \bibinfo {author} {\bibfnamefont {A.~R.}\
  \bibnamefont {Pol}}, \bibinfo {author} {\bibfnamefont {E.~R.}\ \bibnamefont
  {Morales}}, \bibinfo {author} {\bibfnamefont {M.}~\bibnamefont
  {Sakellariadou}}, \bibinfo {author} {\bibfnamefont {I.~D.}\ \bibnamefont
  {Saltas}}, \bibinfo {author} {\bibfnamefont {M.}~\bibnamefont {Scalisi}},
  \bibinfo {author} {\bibfnamefont {K.}~\bibnamefont {Schmitz}}, \bibinfo
  {author} {\bibfnamefont {P.}~\bibnamefont {Schwaller}}, \bibinfo {author}
  {\bibfnamefont {O.}~\bibnamefont {Sergijenko}}, \bibinfo {author}
  {\bibfnamefont {G.}~\bibnamefont {Servant}}, \bibinfo {author} {\bibfnamefont
  {P.}~\bibnamefont {Simakachorn}}, \bibinfo {author} {\bibfnamefont
  {L.}~\bibnamefont {Sorbo}}, \bibinfo {author} {\bibfnamefont
  {L.}~\bibnamefont {Sousa}}, \bibinfo {author} {\bibfnamefont
  {L.}~\bibnamefont {Speri}}, \bibinfo {author} {\bibfnamefont {D.~A.}\
  \bibnamefont {Steer}}, \bibinfo {author} {\bibfnamefont {N.}~\bibnamefont
  {Tamanini}}, \bibinfo {author} {\bibfnamefont {G.}~\bibnamefont {Tasinato}},
  \bibinfo {author} {\bibfnamefont {J.}~\bibnamefont {Torrado}}, \bibinfo
  {author} {\bibfnamefont {C.}~\bibnamefont {Unal}}, \bibinfo {author}
  {\bibfnamefont {V.}~\bibnamefont {Vennin}}, \bibinfo {author} {\bibfnamefont
  {D.}~\bibnamefont {Vernieri}}, \bibinfo {author} {\bibfnamefont
  {F.}~\bibnamefont {Vernizzi}}, \bibinfo {author} {\bibfnamefont
  {M.}~\bibnamefont {Volonteri}}, \bibinfo {author} {\bibfnamefont {J.~M.}\
  \bibnamefont {Wachter}}, \bibinfo {author} {\bibfnamefont {D.}~\bibnamefont
  {Wands}}, \bibinfo {author} {\bibfnamefont {L.~T.}\ \bibnamefont
  {Witkowski}}, \bibinfo {author} {\bibfnamefont {M.}~\bibnamefont
  {Zumalac{\'a}rregui}}, \bibinfo {author} {\bibfnamefont {J.}~\bibnamefont
  {Annis}}, \bibinfo {author} {\bibfnamefont {F.~R.}\ \bibnamefont {Ares}},
  \bibinfo {author} {\bibfnamefont {P.~P.}\ \bibnamefont {Avelino}}, \bibinfo
  {author} {\bibfnamefont {A.}~\bibnamefont {Avgoustidis}}, \bibinfo {author}
  {\bibfnamefont {E.}~\bibnamefont {Barausse}}, \bibinfo {author}
  {\bibfnamefont {A.}~\bibnamefont {Bonilla}}, \bibinfo {author} {\bibfnamefont
  {C.}~\bibnamefont {Bonvin}}, \bibinfo {author} {\bibfnamefont
  {P.}~\bibnamefont {Bosso}}, \bibinfo {author} {\bibfnamefont
  {M.}~\bibnamefont {Calabrese}}, \bibinfo {author} {\bibfnamefont
  {M.}~\bibnamefont {{\c{C}}al{\i}{\c{s}}kan}}, \bibinfo {author}
  {\bibfnamefont {J.~A.~R.}\ \bibnamefont {Cembranos}}, \bibinfo {author}
  {\bibfnamefont {M.}~\bibnamefont {Chala}}, \bibinfo {author} {\bibfnamefont
  {D.}~\bibnamefont {Chernoff}}, \bibinfo {author} {\bibfnamefont
  {K.}~\bibnamefont {Clough}}, \bibinfo {author} {\bibfnamefont
  {A.}~\bibnamefont {Criswell}}, \bibinfo {author} {\bibfnamefont
  {S.}~\bibnamefont {Das}}, \bibinfo {author} {\bibfnamefont {A.~d.}\
  \bibnamefont {Silva}}, \bibinfo {author} {\bibfnamefont {P.}~\bibnamefont
  {Dayal}}, \bibinfo {author} {\bibfnamefont {V.}~\bibnamefont {Domcke}},
  \bibinfo {author} {\bibfnamefont {R.}~\bibnamefont {Durrer}}, \bibinfo
  {author} {\bibfnamefont {R.}~\bibnamefont {Easther}}, \bibinfo {author}
  {\bibfnamefont {S.}~\bibnamefont {Escoffier}}, \bibinfo {author}
  {\bibfnamefont {S.}~\bibnamefont {Ferrans}}, \bibinfo {author} {\bibfnamefont
  {C.}~\bibnamefont {Fryer}}, \bibinfo {author} {\bibfnamefont
  {J.}~\bibnamefont {Gair}}, \bibinfo {author} {\bibfnamefont {C.}~\bibnamefont
  {Gordon}}, \bibinfo {author} {\bibfnamefont {M.}~\bibnamefont {Hendry}},
  \bibinfo {author} {\bibfnamefont {M.}~\bibnamefont {Hindmarsh}}, \bibinfo
  {author} {\bibfnamefont {D.~C.}\ \bibnamefont {Hooper}}, \bibinfo {author}
  {\bibfnamefont {E.}~\bibnamefont {Kajfasz}}, \bibinfo {author} {\bibfnamefont
  {J.}~\bibnamefont {Kopp}}, \bibinfo {author} {\bibfnamefont {S.~M.}\
  \bibnamefont {Koushiappas}}, \bibinfo {author} {\bibfnamefont
  {U.}~\bibnamefont {Kumar}}, \bibinfo {author} {\bibfnamefont
  {M.}~\bibnamefont {Kunz}}, \bibinfo {author} {\bibfnamefont {M.}~\bibnamefont
  {Lagos}}, \bibinfo {author} {\bibfnamefont {M.}~\bibnamefont {Lilley}},
  \bibinfo {author} {\bibfnamefont {J.}~\bibnamefont {Lizarraga}}, \bibinfo
  {author} {\bibfnamefont {F.~S.~N.}\ \bibnamefont {Lobo}}, \bibinfo {author}
  {\bibfnamefont {A.}~\bibnamefont {Maleknejad}}, \bibinfo {author}
  {\bibfnamefont {C.~J. A.~P.}\ \bibnamefont {Martins}}, \bibinfo {author}
  {\bibfnamefont {P.~D.}\ \bibnamefont {Meerburg}}, \bibinfo {author}
  {\bibfnamefont {R.}~\bibnamefont {Meyer}}, \bibinfo {author} {\bibfnamefont
  {J.~P.}\ \bibnamefont {Mimoso}}, \bibinfo {author} {\bibfnamefont
  {S.}~\bibnamefont {Nesseris}}, \bibinfo {author} {\bibfnamefont
  {N.}~\bibnamefont {Nunes}}, \bibinfo {author} {\bibfnamefont
  {V.}~\bibnamefont {Oikonomou}}, \bibinfo {author} {\bibfnamefont
  {G.}~\bibnamefont {Orlando}}, \bibinfo {author} {\bibfnamefont
  {O.}~\bibnamefont {{\"O}zsoy}}, \bibinfo {author} {\bibfnamefont
  {F.}~\bibnamefont {Pacucci}}, \bibinfo {author} {\bibfnamefont
  {A.}~\bibnamefont {Palmese}}, \bibinfo {author} {\bibfnamefont
  {A.}~\bibnamefont {Petiteau}}, \bibinfo {author} {\bibfnamefont
  {L.}~\bibnamefont {Pinol}}, \bibinfo {author} {\bibfnamefont {S.~P.}\
  \bibnamefont {Zwart}}, \bibinfo {author} {\bibfnamefont {G.}~\bibnamefont
  {Pratten}}, \bibinfo {author} {\bibfnamefont {T.}~\bibnamefont {Prokopec}},
  \bibinfo {author} {\bibfnamefont {J.}~\bibnamefont {Quenby}}, \bibinfo
  {author} {\bibfnamefont {S.}~\bibnamefont {Rastgoo}}, \bibinfo {author}
  {\bibfnamefont {D.}~\bibnamefont {Roest}}, \bibinfo {author} {\bibfnamefont
  {K.}~\bibnamefont {Rummukainen}}, \bibinfo {author} {\bibfnamefont
  {C.}~\bibnamefont {Schimd}}, \bibinfo {author} {\bibfnamefont
  {A.}~\bibnamefont {Secroun}}, \bibinfo {author} {\bibfnamefont
  {A.}~\bibnamefont {Sesana}}, \bibinfo {author} {\bibfnamefont {C.~F.}\
  \bibnamefont {Sopuerta}}, \bibinfo {author} {\bibfnamefont {I.}~\bibnamefont
  {Tereno}}, \bibinfo {author} {\bibfnamefont {A.}~\bibnamefont {Tolley}},
  \bibinfo {author} {\bibfnamefont {J.}~\bibnamefont {Urrestilla}}, \bibinfo
  {author} {\bibfnamefont {E.~C.}\ \bibnamefont {Vagenas}}, \bibinfo {author}
  {\bibfnamefont {J.}~\bibnamefont {van~de Vis}}, \bibinfo {author}
  {\bibfnamefont {R.}~\bibnamefont {van~de Weygaert}}, \bibinfo {author}
  {\bibfnamefont {B.}~\bibnamefont {Wardell}}, \bibinfo {author} {\bibfnamefont
  {D.~J.}\ \bibnamefont {Weir}}, \bibinfo {author} {\bibfnamefont
  {G.}~\bibnamefont {White}}, \bibinfo {author} {\bibfnamefont
  {B.}~\bibnamefont {{\'{S}}wie{\.{z}}ewska}}, \bibinfo {author} {\bibfnamefont
  {V.~I.}\ \bibnamefont {Zhdanov}},\ and\ \bibinfo {author} {\bibfnamefont
  {T.~L. C.~W.}\ \bibnamefont {Group}},\ }\bibfield  {title} {\bibinfo {title}
  {Cosmology with the laser interferometer space antenna},\ }\href
  {https://doi.org/10.1007/s41114-023-00045-2} {\bibfield  {journal} {\bibinfo
  {journal} {Living Reviews in Relativity}\ }\textbf {\bibinfo {volume} {26}},\
  \bibinfo {pages} {5} (\bibinfo {year} {2023})}\BibitemShut {NoStop}%
\bibitem [{rep()}]{repo}%
  \BibitemOpen
  \href@noop {} {}\bibinfo {howpublished}
  {\href{https://github.com/jpmvferreira/testing-L-free-fQ-cosmology}{https://github.com/jpmvferreira/testing-L-free-fQ-cosmology}}\BibitemShut
  {NoStop}%
\bibitem [{\citenamefont {J\"arv}\ \emph {et~al.}(2018)\citenamefont {J\"arv},
  \citenamefont {R\"unkla}, \citenamefont {Saal},\ and\ \citenamefont
  {Vilson}}]{Jaerv2018}%
  \BibitemOpen
  \bibfield  {author} {\bibinfo {author} {\bibfnamefont {L.}~\bibnamefont
  {J\"arv}}, \bibinfo {author} {\bibfnamefont {M.}~\bibnamefont {R\"unkla}},
  \bibinfo {author} {\bibfnamefont {M.}~\bibnamefont {Saal}},\ and\ \bibinfo
  {author} {\bibfnamefont {O.}~\bibnamefont {Vilson}},\ }\bibfield  {title}
  {\bibinfo {title} {{Nonmetricity formulation of general relativity and its
  scalar-tensor extension}},\ }\href
  {https://doi.org/10.1103/PhysRevD.97.124025} {\bibfield  {journal} {\bibinfo
  {journal} {Phys. Rev. D}\ }\textbf {\bibinfo {volume} {97}},\ \bibinfo
  {pages} {124025} (\bibinfo {year} {2018})},\ \Eprint
  {https://arxiv.org/abs/1802.00492} {arXiv:1802.00492 [gr-qc]} \BibitemShut
  {NoStop}%
\bibitem [{\citenamefont {Jiménez}\ \emph {et~al.}(2019)\citenamefont
  {Jiménez}, \citenamefont {Heisenberg}, \citenamefont {Koivisto},\ and\
  \citenamefont {Pekar}}]{Jimenez2019}%
  \BibitemOpen
  \bibfield  {author} {\bibinfo {author} {\bibfnamefont {J.~B.}\ \bibnamefont
  {Jiménez}}, \bibinfo {author} {\bibfnamefont {L.}~\bibnamefont
  {Heisenberg}}, \bibinfo {author} {\bibfnamefont {T.~S.}\ \bibnamefont
  {Koivisto}},\ and\ \bibinfo {author} {\bibfnamefont {S.}~\bibnamefont
  {Pekar}},\ }\bibfield  {title} {\bibinfo {title} {Cosmology in $f({Q})$
  geometry},\ }\bibfield  {journal} {\bibinfo  {journal} {Phys. Rev. D 101,
  103507 (2020)}\ }\href {https://doi.org/10.1103/PhysRevD.101.103507}
  {10.1103/PhysRevD.101.103507} (\bibinfo {year} {2019}),\ \Eprint
  {https://arxiv.org/abs/1906.10027} {arXiv:1906.10027 [gr-qc]} \BibitemShut
  {NoStop}%
\bibitem [{\citenamefont {Jim\'enez}\ \emph {et~al.}(2018)\citenamefont
  {Jim\'enez}, \citenamefont {Heisenberg},\ and\ \citenamefont
  {Koivisto}}]{Jimenez2018}%
  \BibitemOpen
  \bibfield  {author} {\bibinfo {author} {\bibfnamefont {J.}~\bibnamefont
  {Jim\'enez}}, \bibinfo {author} {\bibfnamefont {L.}~\bibnamefont
  {Heisenberg}},\ and\ \bibinfo {author} {\bibfnamefont {T.~S.}\ \bibnamefont
  {Koivisto}},\ }\bibfield  {title} {\bibinfo {title} {Teleparallel palatini
  theories},\ }\href {https://doi.org/10.1088/1475-7516/2018/08/039} {\bibfield
   {journal} {\bibinfo  {journal} {Journal of Cosmology and Astroparticle
  Physics}\ }\textbf {\bibinfo {volume} {08}}\bibfield  {number} {\bibinfo
  {number} { (08)},\ \bibinfo {pages} {039}},\ }\Eprint
  {https://arxiv.org/abs/1803.10185} {arXiv:1803.10185 [gr-qc]} \BibitemShut
  {NoStop}%
\bibitem [{\citenamefont {Koivisto}(2006)}]{Koivisto2005}%
  \BibitemOpen
  \bibfield  {author} {\bibinfo {author} {\bibfnamefont {T.}~\bibnamefont
  {Koivisto}},\ }\bibfield  {title} {\bibinfo {title} {A note on covariant
  conservation of energy–momentum in modified gravities},\ }\href
  {https://doi.org/10.1088/0264-9381/23/12/N01} {\bibfield  {journal} {\bibinfo
   {journal} {Classical and Quantum Gravity}\ }\textbf {\bibinfo {volume}
  {23}},\ \bibinfo {pages} {4289} (\bibinfo {year} {2006})}\BibitemShut
  {NoStop}%
\bibitem [{\citenamefont {Anagnostopoulos}\ \emph {et~al.}(2021)\citenamefont
  {Anagnostopoulos}, \citenamefont {Basilakos},\ and\ \citenamefont
  {Saridakis}}]{Anagnostopoulos2021}%
  \BibitemOpen
  \bibfield  {author} {\bibinfo {author} {\bibfnamefont {F.~K.}\ \bibnamefont
  {Anagnostopoulos}}, \bibinfo {author} {\bibfnamefont {S.}~\bibnamefont
  {Basilakos}},\ and\ \bibinfo {author} {\bibfnamefont {E.~N.}\ \bibnamefont
  {Saridakis}},\ }\bibfield  {title} {\bibinfo {title} {First evidence that
  non-metricity f({Q}) gravity could challenge ${\Lambda}${CDM}},\ }\href
  {https://doi.org/https://doi.org/10.1016/j.physletb.2021.136634} {\bibfield
  {journal} {\bibinfo  {journal} {Physics Letters B}\ }\textbf {\bibinfo
  {volume} {822}},\ \bibinfo {pages} {136634} (\bibinfo {year}
  {2021})}\BibitemShut {NoStop}%
\bibitem [{\citenamefont {Anagnostopoulos}\ \emph {et~al.}(2023)\citenamefont
  {Anagnostopoulos}, \citenamefont {Gakis}, \citenamefont {Saridakis},\ and\
  \citenamefont {Basilakos}}]{Anagnostopoulos2022}%
  \BibitemOpen
  \bibfield  {author} {\bibinfo {author} {\bibfnamefont {F.~K.}\ \bibnamefont
  {Anagnostopoulos}}, \bibinfo {author} {\bibfnamefont {V.}~\bibnamefont
  {Gakis}}, \bibinfo {author} {\bibfnamefont {E.~N.}\ \bibnamefont
  {Saridakis}},\ and\ \bibinfo {author} {\bibfnamefont {S.}~\bibnamefont
  {Basilakos}},\ }\bibfield  {title} {\bibinfo {title} {New models and big bang
  nucleosynthesis constraints in f({Q}) gravity},\ }\href
  {https://doi.org/10.1140/epjc/s10052-023-11190-x} {\bibfield  {journal}
  {\bibinfo  {journal} {The European Physical Journal C}\ }\textbf {\bibinfo
  {volume} {83}},\ \bibinfo {pages} {58} (\bibinfo {year} {2023})}\BibitemShut
  {NoStop}%
\bibitem [{\citenamefont {Bahamonde}\ \emph {et~al.}(2018)\citenamefont
  {Bahamonde}, \citenamefont {Böhmer}, \citenamefont {Carloni}, \citenamefont
  {Copeland}, \citenamefont {Fang},\ and\ \citenamefont
  {Tamanini}}]{Bahamonde2018}%
  \BibitemOpen
  \bibfield  {author} {\bibinfo {author} {\bibfnamefont {S.}~\bibnamefont
  {Bahamonde}}, \bibinfo {author} {\bibfnamefont {C.~G.}\ \bibnamefont
  {Böhmer}}, \bibinfo {author} {\bibfnamefont {S.}~\bibnamefont {Carloni}},
  \bibinfo {author} {\bibfnamefont {E.~J.}\ \bibnamefont {Copeland}}, \bibinfo
  {author} {\bibfnamefont {W.}~\bibnamefont {Fang}},\ and\ \bibinfo {author}
  {\bibfnamefont {N.}~\bibnamefont {Tamanini}},\ }\bibfield  {title} {\bibinfo
  {title} {Dynamical systems applied to cosmology: Dark energy and modified
  gravity},\ }\href
  {https://doi.org/https://doi.org/10.1016/j.physrep.2018.09.001} {\bibfield
  {journal} {\bibinfo  {journal} {Physics Reports}\ }\textbf {\bibinfo {volume}
  {775-777}},\ \bibinfo {pages} {1} (\bibinfo {year} {2018})},\ \bibinfo {note}
  {dynamical systems applied to cosmology: Dark energy and modified
  gravity}\BibitemShut {NoStop}%
\bibitem [{\citenamefont {Scolnic}\ \emph {et~al.}(2018)\citenamefont
  {Scolnic}, \citenamefont {Jones}, \citenamefont {Rest}, \citenamefont {Pan},
  \citenamefont {Chornock}, \citenamefont {Foley}, \citenamefont {Huber},
  \citenamefont {Kessler}, \citenamefont {Narayan}, \citenamefont {Riess},
  \citenamefont {Rodney}, \citenamefont {Berger}, \citenamefont {Brout},
  \citenamefont {Challis}, \citenamefont {Drout}, \citenamefont {Finkbeiner},
  \citenamefont {Lunnan}, \citenamefont {Kirshner}, \citenamefont {Sanders},
  \citenamefont {Schlafly}, \citenamefont {Smartt}, \citenamefont {Stubbs},
  \citenamefont {Tonry}, \citenamefont {Wood-Vasey}, \citenamefont {Foley},
  \citenamefont {Hand}, \citenamefont {Johnson}, \citenamefont {Burgett},
  \citenamefont {Chambers}, \citenamefont {Draper}, \citenamefont {Hodapp},
  \citenamefont {Kaiser}, \citenamefont {Kudritzki}, \citenamefont {Magnier},
  \citenamefont {Metcalfe}, \citenamefont {Bresolin}, \citenamefont {Gall},
  \citenamefont {Kotak}, \citenamefont {McCrum},\ and\ \citenamefont
  {Smith}}]{pantheon}%
  \BibitemOpen
  \bibfield  {author} {\bibinfo {author} {\bibfnamefont {D.~M.}\ \bibnamefont
  {Scolnic}}, \bibinfo {author} {\bibfnamefont {D.~O.}\ \bibnamefont {Jones}},
  \bibinfo {author} {\bibfnamefont {A.}~\bibnamefont {Rest}}, \bibinfo {author}
  {\bibfnamefont {Y.~C.}\ \bibnamefont {Pan}}, \bibinfo {author} {\bibfnamefont
  {R.}~\bibnamefont {Chornock}}, \bibinfo {author} {\bibfnamefont {R.~J.}\
  \bibnamefont {Foley}}, \bibinfo {author} {\bibfnamefont {M.~E.}\ \bibnamefont
  {Huber}}, \bibinfo {author} {\bibfnamefont {R.}~\bibnamefont {Kessler}},
  \bibinfo {author} {\bibfnamefont {G.}~\bibnamefont {Narayan}}, \bibinfo
  {author} {\bibfnamefont {A.~G.}\ \bibnamefont {Riess}}, \bibinfo {author}
  {\bibfnamefont {S.}~\bibnamefont {Rodney}}, \bibinfo {author} {\bibfnamefont
  {E.}~\bibnamefont {Berger}}, \bibinfo {author} {\bibfnamefont {D.~J.}\
  \bibnamefont {Brout}}, \bibinfo {author} {\bibfnamefont {P.~J.}\ \bibnamefont
  {Challis}}, \bibinfo {author} {\bibfnamefont {M.}~\bibnamefont {Drout}},
  \bibinfo {author} {\bibfnamefont {D.}~\bibnamefont {Finkbeiner}}, \bibinfo
  {author} {\bibfnamefont {R.}~\bibnamefont {Lunnan}}, \bibinfo {author}
  {\bibfnamefont {R.~P.}\ \bibnamefont {Kirshner}}, \bibinfo {author}
  {\bibfnamefont {N.~E.}\ \bibnamefont {Sanders}}, \bibinfo {author}
  {\bibfnamefont {E.}~\bibnamefont {Schlafly}}, \bibinfo {author}
  {\bibfnamefont {S.}~\bibnamefont {Smartt}}, \bibinfo {author} {\bibfnamefont
  {C.~W.}\ \bibnamefont {Stubbs}}, \bibinfo {author} {\bibfnamefont
  {J.}~\bibnamefont {Tonry}}, \bibinfo {author} {\bibfnamefont {W.~M.}\
  \bibnamefont {Wood-Vasey}}, \bibinfo {author} {\bibfnamefont
  {M.}~\bibnamefont {Foley}}, \bibinfo {author} {\bibfnamefont
  {J.}~\bibnamefont {Hand}}, \bibinfo {author} {\bibfnamefont {E.}~\bibnamefont
  {Johnson}}, \bibinfo {author} {\bibfnamefont {W.~S.}\ \bibnamefont
  {Burgett}}, \bibinfo {author} {\bibfnamefont {K.~C.}\ \bibnamefont
  {Chambers}}, \bibinfo {author} {\bibfnamefont {P.~W.}\ \bibnamefont
  {Draper}}, \bibinfo {author} {\bibfnamefont {K.~W.}\ \bibnamefont {Hodapp}},
  \bibinfo {author} {\bibfnamefont {N.}~\bibnamefont {Kaiser}}, \bibinfo
  {author} {\bibfnamefont {R.~P.}\ \bibnamefont {Kudritzki}}, \bibinfo {author}
  {\bibfnamefont {E.~A.}\ \bibnamefont {Magnier}}, \bibinfo {author}
  {\bibfnamefont {N.}~\bibnamefont {Metcalfe}}, \bibinfo {author}
  {\bibfnamefont {F.}~\bibnamefont {Bresolin}}, \bibinfo {author}
  {\bibfnamefont {E.}~\bibnamefont {Gall}}, \bibinfo {author} {\bibfnamefont
  {R.}~\bibnamefont {Kotak}}, \bibinfo {author} {\bibfnamefont
  {M.}~\bibnamefont {McCrum}},\ and\ \bibinfo {author} {\bibfnamefont {K.~W.}\
  \bibnamefont {Smith}} (\bibinfo {collaboration} {Pan-STARRS1}),\ }\bibfield
  {title} {\bibinfo {title} {The complete light-curve sample of
  spectroscopically confirmed {T}ype {I}a {S}upernovae from {Pan-STARRS1} and
  cosmological constraints from the combined {P}antheon sample},\ }\href
  {https://doi.org/10.3847/1538-4357/aab9bb} {\bibfield  {journal} {\bibinfo
  {journal} {Astrophys. J.}\ }\textbf {\bibinfo {volume} {859}},\ \bibinfo
  {pages} {101} (\bibinfo {year} {2018})},\ \Eprint
  {https://arxiv.org/abs/1710.00845} {arXiv:1710.00845 [astro-ph.CO]}
  \BibitemShut {NoStop}%
\bibitem [{pan()}]{pantheon-repo}%
  \BibitemOpen
  \href@noop {} {}\bibinfo {howpublished}
  {\href{https://github.com/dscolnic/Pantheon}{https://github.com/dscolnic/Pantheon}}\BibitemShut
  {NoStop}%
\bibitem [{\citenamefont {Goliath}\ \emph {et~al.}(2001)\citenamefont
  {Goliath}, \citenamefont {Amanullah}, \citenamefont {Astier}, \citenamefont
  {Goobar},\ and\ \citenamefont {Pain}}]{Goliath2001}%
  \BibitemOpen
  \bibfield  {author} {\bibinfo {author} {\bibfnamefont {M.}~\bibnamefont
  {Goliath}}, \bibinfo {author} {\bibfnamefont {R.}~\bibnamefont {Amanullah}},
  \bibinfo {author} {\bibfnamefont {P.}~\bibnamefont {Astier}}, \bibinfo
  {author} {\bibfnamefont {A.}~\bibnamefont {Goobar}},\ and\ \bibinfo {author}
  {\bibfnamefont {R.}~\bibnamefont {Pain}},\ }\bibfield  {title} {\bibinfo
  {title} {Supernovae and the nature of the dark energy},\ }\href
  {https://doi.org/10.1051/0004-6361:20011398} {\bibfield  {journal} {\bibinfo
  {journal} {Astronomy {\&} Astrophysics}\ }\textbf {\bibinfo {volume} {380}},\
  \bibinfo {pages} {6} (\bibinfo {year} {2001})},\ \Eprint
  {https://arxiv.org/abs/astro-ph/0104009} {arXiv:astro-ph/0104009 [astro-ph]}
  \BibitemShut {NoStop}%
\bibitem [{\citenamefont {Zhai}\ and\ \citenamefont {Wang}(2019)}]{Zhai2018}%
  \BibitemOpen
  \bibfield  {author} {\bibinfo {author} {\bibfnamefont {Z.}~\bibnamefont
  {Zhai}}\ and\ \bibinfo {author} {\bibfnamefont {Y.}~\bibnamefont {Wang}},\
  }\bibfield  {title} {\bibinfo {title} {{Robust and model-independent
  cosmological constraints from distance measurements}},\ }\href
  {https://doi.org/10.1088/1475-7516/2019/07/005} {\bibfield  {journal}
  {\bibinfo  {journal} {JCAP}\ }\textbf {\bibinfo {volume} {07}},\ \bibinfo
  {pages} {005}},\ \Eprint {https://arxiv.org/abs/1811.07425} {arXiv:1811.07425
  [astro-ph.CO]} \BibitemShut {NoStop}%
\bibitem [{\citenamefont {Hu}\ and\ \citenamefont {Sugiyama}(1996)}]{Hu1995}%
  \BibitemOpen
  \bibfield  {author} {\bibinfo {author} {\bibfnamefont {W.}~\bibnamefont
  {Hu}}\ and\ \bibinfo {author} {\bibfnamefont {N.}~\bibnamefont {Sugiyama}},\
  }\bibfield  {title} {\bibinfo {title} {Small-scale cosmological
  perturbations: An analytic approach},\ }\href
  {https://doi.org/10.1086/177989} {\bibfield  {journal} {\bibinfo  {journal}
  {The Astrophysical Journal}\ }\textbf {\bibinfo {volume} {471}},\ \bibinfo
  {pages} {542} (\bibinfo {year} {1996})}\BibitemShut {NoStop}%
\bibitem [{\citenamefont {Perez}\ \emph {et~al.}(2023)\citenamefont {Perez},
  \citenamefont {Peracaula},\ and\ \citenamefont {Singh}}]{Perez2023}%
  \BibitemOpen
  \bibfield  {author} {\bibinfo {author} {\bibfnamefont {J.~d.~C.}\
  \bibnamefont {Perez}}, \bibinfo {author} {\bibfnamefont {J.~S.}\ \bibnamefont
  {Peracaula}},\ and\ \bibinfo {author} {\bibfnamefont {C.~P.}\ \bibnamefont
  {Singh}},\ }\bibfield  {title} {\bibinfo {title} {Running vacuum in
  brans-dicke theory: a possible cure for the $\sigma_8$ and ${H}_0$
  tensions},\ }\href@noop {} {\  (\bibinfo {year} {2023})},\ \Eprint
  {https://arxiv.org/abs/2302.04807} {arXiv:2302.04807 [astro-ph.CO]}
  \BibitemShut {NoStop}%
\bibitem [{\citenamefont {Collaboration}(2020)}]{Planck2018}%
  \BibitemOpen
  \bibfield  {author} {\bibinfo {author} {\bibfnamefont {P.}~\bibnamefont
  {Collaboration}},\ }\bibfield  {title} {\bibinfo {title} {Planck 2018
  results},\ }\href {https://doi.org/10.1051/0004-6361/201833910} {\bibfield
  {journal} {\bibinfo  {journal} {Astronomy {\&} Astrophysics}\ }\textbf
  {\bibinfo {volume} {641}},\ \bibinfo {pages} {A6} (\bibinfo {year}
  {2020})}\BibitemShut {NoStop}%
\bibitem [{\citenamefont {Ferreira}\ \emph {et~al.}(2022)\citenamefont
  {Ferreira}, \citenamefont {Barreiro}, \citenamefont {Mimoso},\ and\
  \citenamefont {Nunes}}]{Ferreira2022}%
  \BibitemOpen
  \bibfield  {author} {\bibinfo {author} {\bibfnamefont {J.}~\bibnamefont
  {Ferreira}}, \bibinfo {author} {\bibfnamefont {T.}~\bibnamefont {Barreiro}},
  \bibinfo {author} {\bibfnamefont {J.}~\bibnamefont {Mimoso}},\ and\ \bibinfo
  {author} {\bibfnamefont {N.~J.}\ \bibnamefont {Nunes}},\ }\bibfield  {title}
  {\bibinfo {title} {Forecasting {F(Q)} cosmology with {$\Lambda$CDM}
  background using standard sirens},\ }\href
  {https://doi.org/10.1103/physrevd.105.123531} {\bibfield  {journal} {\bibinfo
   {journal} {Phys. Rev. D}\ }\textbf {\bibinfo {volume} {105}},\ \bibinfo
  {pages} {123531} (\bibinfo {year} {2022})},\ \Eprint
  {https://arxiv.org/abs/2203.13788} {arXiv:2203.13788 [astro-ph.CO]}
  \BibitemShut {NoStop}%
\bibitem [{\citenamefont {Riddell}\ \emph {et~al.}(2021)\citenamefont
  {Riddell}, \citenamefont {Hartikainen},\ and\ \citenamefont
  {Carter}}]{PyStan}%
  \BibitemOpen
  \bibfield  {author} {\bibinfo {author} {\bibfnamefont {A.}~\bibnamefont
  {Riddell}}, \bibinfo {author} {\bibfnamefont {A.}~\bibnamefont
  {Hartikainen}},\ and\ \bibinfo {author} {\bibfnamefont {M.}~\bibnamefont
  {Carter}},\ }\href@noop {} {\bibinfo {title} {pystan (3.0.0)}},\ \bibinfo
  {howpublished} {PyPi} (\bibinfo {year} {2021})\BibitemShut {NoStop}%
\bibitem [{\citenamefont {Team}(2021)}]{Stan}%
  \BibitemOpen
  \bibfield  {author} {\bibinfo {author} {\bibfnamefont {S.~D.}\ \bibnamefont
  {Team}},\ }\href {https://mc-stan.org} {\emph {\bibinfo {title} {Stan
  Modeling Language Users Guide and Reference Manual}}},\ \bibinfo {edition}
  {2nd}\ ed. (\bibinfo {year} {2021})\BibitemShut {NoStop}%
\bibitem [{\citenamefont {Lewis}(2019)}]{GetDist}%
  \BibitemOpen
  \bibfield  {author} {\bibinfo {author} {\bibfnamefont {A.}~\bibnamefont
  {Lewis}},\ }\bibfield  {title} {\bibinfo {title} {Getdist: a python package
  for analysing monte carlo samples},\ }\href@noop {} {\  (\bibinfo {year}
  {2019})},\ \Eprint {https://arxiv.org/abs/1910.13970} {arXiv:1910.13970
  [astro-ph.IM]} \BibitemShut {NoStop}%
\bibitem [{\citenamefont {Kumar}\ \emph {et~al.}(2019)\citenamefont {Kumar},
  \citenamefont {Carroll}, \citenamefont {Hartikainen},\ and\ \citenamefont
  {Martin}}]{arviz}%
  \BibitemOpen
  \bibfield  {author} {\bibinfo {author} {\bibfnamefont {R.}~\bibnamefont
  {Kumar}}, \bibinfo {author} {\bibfnamefont {C.}~\bibnamefont {Carroll}},
  \bibinfo {author} {\bibfnamefont {A.}~\bibnamefont {Hartikainen}},\ and\
  \bibinfo {author} {\bibfnamefont {O.}~\bibnamefont {Martin}},\ }\bibfield
  {title} {\bibinfo {title} {{ArviZ} a unified library for exploratory analysis
  of bayesian models in python},\ }\href {https://doi.org/10.21105/joss.01143}
  {\bibfield  {journal} {\bibinfo  {journal} {Journal of Open Source Software}\
  }\textbf {\bibinfo {volume} {4}},\ \bibinfo {pages} {1143} (\bibinfo {year}
  {2019})}\BibitemShut {NoStop}%
\bibitem [{\citenamefont {Vehtari}\ \emph {et~al.}(2021)\citenamefont
  {Vehtari}, \citenamefont {Gelman}, \citenamefont {Simpson}, \citenamefont
  {Carpenter},\ and\ \citenamefont {Bürkner}}]{Vehtari2021}%
  \BibitemOpen
  \bibfield  {author} {\bibinfo {author} {\bibfnamefont {A.}~\bibnamefont
  {Vehtari}}, \bibinfo {author} {\bibfnamefont {A.}~\bibnamefont {Gelman}},
  \bibinfo {author} {\bibfnamefont {D.}~\bibnamefont {Simpson}}, \bibinfo
  {author} {\bibfnamefont {B.}~\bibnamefont {Carpenter}},\ and\ \bibinfo
  {author} {\bibfnamefont {P.-C.}\ \bibnamefont {Bürkner}},\ }\bibfield
  {title} {\bibinfo {title} {Rank-normalization, folding, and localization: An
  improved {R}ˆ for assessing convergence of {MCMC} (with discussion)},\
  }\bibfield  {journal} {\bibinfo  {journal} {Bayesian Analysis}\ }\textbf
  {\bibinfo {volume} {16}},\ \href {https://doi.org/10.1214/20-ba1221}
  {10.1214/20-ba1221} (\bibinfo {year} {2021})\BibitemShut {NoStop}%
\bibitem [{\citenamefont {Nunes}\ \emph {et~al.}(2017)\citenamefont {Nunes},
  \citenamefont {Pan}, \citenamefont {Saridakis},\ and\ \citenamefont
  {Abreu}}]{1610.07518}%
  \BibitemOpen
  \bibfield  {author} {\bibinfo {author} {\bibfnamefont {R.~C.}\ \bibnamefont
  {Nunes}}, \bibinfo {author} {\bibfnamefont {S.}~\bibnamefont {Pan}}, \bibinfo
  {author} {\bibfnamefont {E.~N.}\ \bibnamefont {Saridakis}},\ and\ \bibinfo
  {author} {\bibfnamefont {E.~M.}\ \bibnamefont {Abreu}},\ }\bibfield  {title}
  {\bibinfo {title} {New observational constraints on $f({R})$ gravity from
  cosmic chronometers},\ }\href {https://doi.org/10.1088/1475-7516/2017/01/005}
  {\bibfield  {journal} {\bibinfo  {journal} {Journal of Cosmology and
  Astroparticle Physics}\ }\textbf {\bibinfo {volume} {2017}}\bibinfo  {number}
  { (01)},\ \bibinfo {pages} {005}}\BibitemShut {NoStop}%
\bibitem [{\citenamefont {dos Santos}\ \emph {et~al.}(2022)\citenamefont {dos
  Santos}, \citenamefont {Gonzalez},\ and\ \citenamefont {Silva}}]{2112.15249}%
  \BibitemOpen
\bibfield  {number} {  }\bibfield  {author} {\bibinfo {author} {\bibfnamefont
  {F.~B.~M.}\ \bibnamefont {dos Santos}}, \bibinfo {author} {\bibfnamefont
  {J.~E.}\ \bibnamefont {Gonzalez}},\ and\ \bibinfo {author} {\bibfnamefont
  {R.}~\bibnamefont {Silva}},\ }\bibfield  {title} {\bibinfo {title}
  {Observational constraints on f({T}) gravity from model-independent data},\
  }\bibfield  {journal} {\bibinfo  {journal} {The European Physical Journal C}\
  }\textbf {\bibinfo {volume} {82}},\ \href
  {https://doi.org/10.1140/epjc/s10052-022-10784-1}
  {10.1140/epjc/s10052-022-10784-1} (\bibinfo {year} {2022})\BibitemShut
  {NoStop}%
\bibitem [{\citenamefont {Lazkoz}\ \emph {et~al.}(2019)\citenamefont {Lazkoz},
  \citenamefont {Lobo}, \citenamefont {Ortiz-Baños},\ and\ \citenamefont
  {Salzano}}]{Lazkoz2019}%
  \BibitemOpen
  \bibfield  {author} {\bibinfo {author} {\bibfnamefont {R.}~\bibnamefont
  {Lazkoz}}, \bibinfo {author} {\bibfnamefont {F.~S.~N.}\ \bibnamefont {Lobo}},
  \bibinfo {author} {\bibfnamefont {M.}~\bibnamefont {Ortiz-Baños}},\ and\
  \bibinfo {author} {\bibfnamefont {V.}~\bibnamefont {Salzano}},\ }\bibfield
  {title} {\bibinfo {title} {Observational constraints of $f({Q})$ gravity},\
  }\href {https://doi.org/10.1103/physrevd.100.104027} {\bibfield  {journal}
  {\bibinfo  {journal} {Physical Review D}\ }\textbf {\bibinfo {volume}
  {100}},\ \bibinfo {pages} {104027} (\bibinfo {year} {2019})},\ \Eprint
  {https://arxiv.org/abs/1907.13219} {arXiv:1907.13219 [gr-qc]} \BibitemShut
  {NoStop}%
\end{thebibliography}%

\end{document}